%% file: lieb_kondo_sc_RajModV21.tex
\documentclass[aps,prb,twocolumn,superscriptaddress,showpacs,preprintnumbers,bibliography]{revtex4-2}
\usepackage[colorlinks,bookmarks=false,citecolor=blue,linkcolor=red,urlcolor=blue]{hyperref}
\usepackage{tikz}
\usetikzlibrary{arrows.meta,decorations.pathreplacing}
\usepackage{xcolor}
\usepackage{booktabs,array}
\definecolor{NavyInk}{HTML}{33338C}
\definecolor{Crit}{HTML}{D85A30}
\definecolor{SoftBlue}{HTML}{E8E8F4}
\definecolor{GreyRule}{HTML}{9A9AA2}
\usepackage{amsmath}

\usepackage{physics}
\usepackage{colortbl,amsthm,txfonts}
\usepackage{verbatim}
\usepackage{graphicx}
\usepackage{epsfig}
\usepackage{dcolumn}

\usepackage{bm}

\usepackage{amsmath,amssymb}
\newcommand{\kv}{\bm{k}}
\newcommand{\qv}{\bm{q}}

\newcommand{\Mset}{\mathcal{M}}

\usepackage{appendix}

\input{prb_zero_modes_setup}
\begin{document}

\title{Lifshitz transition and magnetic superconductivity in flat band systems} 
\author{Sudeshna Pal}
\affiliation{Department of Physics, Indian Institute of Technology, Madras, Chennai 600036, India.}
\author{K.B. Yogendra}
\affiliation{Division of Physics and Applied Physics, School of Physical and Mathematical Sciences,
Nanyang Technological University, Singapore 637371.}
\author{Madhuparna Karmakar}
\email{madhuparna.k@gmail.com}
\affiliation{Department of Physics and Nanotechnology, SRM Institute of Science and Technology,   \\ 
Kattankulathur, Chennai 603203, India}
\author{Rajesh Narayanan}
\email{rnarayanan@zmail.iitm.ac.in}
\affiliation{Department of Physics, Indian Institute of Technology, Madras, Chennai 600036, India.}
\begin{abstract}

A Lifshitz transition is a class of non-symmetry breaking transitions typically associated with metallic systems wherein the Fermi
surface changes its connectivity as the chemical potential sweeps through a
band extremum or a saddle point. The phenomenology naturally adapts itself to the superconducting state, where the object that
reconstructs is the manifold of minimum quasiparticle energy rather than the Fermi surface. This paper concerns Lifshitz-type transitions in multi-band superconductors. Thus, we focus on the attractive Hubbard--Kondo model on the two-dimensional Lieb lattice, with classical core spins coupled to the itinerant electrons, by laying recourse to a real-space Bogoliubov--de Gennes mean-field theory and an analytic Green's function calculation. The competition between the Kondo and pairing channels yields a magnetic superconductor throughout, with $(\pi,\pi)$ order near half filling giving way to $(0,\pi)$ and then, at stronger Kondo coupling, to an incommensurate spiral $(0,q)$. In such a multiband superconductor hosting a flat band, we identify two inequivalent Lifshitz-type transitions. With the core spins decoupled (the Kondo coupling set to zero), the minimum energy contour deforms with chemical potential, a hole pocket about $M$ opening into a perfectly nested square at $\mu=-2t$ and reconnecting into pockets about $\Gamma$; this nesting fixes where $(\pi,\pi)$ order sets in. At finite Kondo coupling, for certain chemical potential,  the dimension of the manifold itself drops. We obtain it analytically for each magnetic order and show that only the spiral carries the minimum energy through zero, nucleating nodes and leaving a gapless magnetic superconductor, with clear signatures in the density of states and the spectral function.
\end{abstract}
\date{\today}
\maketitle

\section{Introduction}
In a seminal paper Lifshitz \cite{lifshitz1960}, proved the 
existence of a class of non-symmetry breaking quantum phase transitions predicated on 
the change in the topology of Fermi surface in metals as a function of the electronic occupancy. 
More specifically, as a function of a tuning parameter like chemical potential, (controlled, for instance via doping), through 
a band extremum or a saddle point, the Fermi surface reconstructs via the appearance or disappearance of a
pocket or through the disruption of a neck connecting adjacent sheets~\cite{lifshitz1960,blanter1994}. Dubbed by the moniker, electronic topological 
transition, or equivalently called the Lifshitz transition, this zero-temperature transition is signalled by a non-analyticity of the thermodynamic potential, and 
gives rise to a dimension-dependent singular contribution to the Density of states (DoS): In 2-d for instance a saddle point contributes a logarithmic 
divergence ~\cite{vanhove1953,markiewicz1997}, whereas a band edge a step discontinuity. This further 
implies that any physical quantity that derives itself from the Fermi surface will 
show anomalous behavior as the system goes through the transition ~\cite{varlamov1989}. Experimentally, such transitions have now been seen in noble metals \cite{blanter1994}, 
iron based pnictides\cite{liu2010}, and also in the normal state of cuprate superconductors.

The Lifshitz phenomenology finds a natural extension from the topology associated with the Fermi surface to that of the zero-energy manifold of 
 the superconductor. For a conventional $s$-wave superconductor, the hard superconducting gap essentially rules out the existence of 
 gapless excitations or a locus of vanishing quasiparticle energy. However, in the presence of competing interactions such as a Zeeman field or 
 magnetic interactions that lead to finite-wave-vector ordering, the minimum quasi-particle energy is driven through zero, thus nucleating nodes on the Fermi surface. 
In other words, just like the Lifshitz transition in normal metals, the topology of such nodal regions can be altered via a suitable control parameter giving rise to a non-symmetry breaking quantum phase transition. Once again, the transition is bereft of a local order-parameter description and just like in the case of metallic systems is signalled by the appearance of non-analyticities in the low energy DoS.

The experimental realization of such transitions is by now well
established, the chemical potential being tuned through a critical
energy by chemical substitution, by hydrostatic or uniaxial pressure,
or by electrostatic gating. In the iron based superconductors a doping
driven Lifshitz transition has been identified through angle-resolved
photoemission and found to coincide with the onset of
superconductivity~\cite{liu2010}, whilst a reconstruction of the Fermi
surface topology has been argued to accompany the evolution of the
underdoped cuprates~\cite{norman2010}.

\begin{figure}
\begin{center}
\includegraphics[height=4.3cm,width=8.2cm,angle=0]{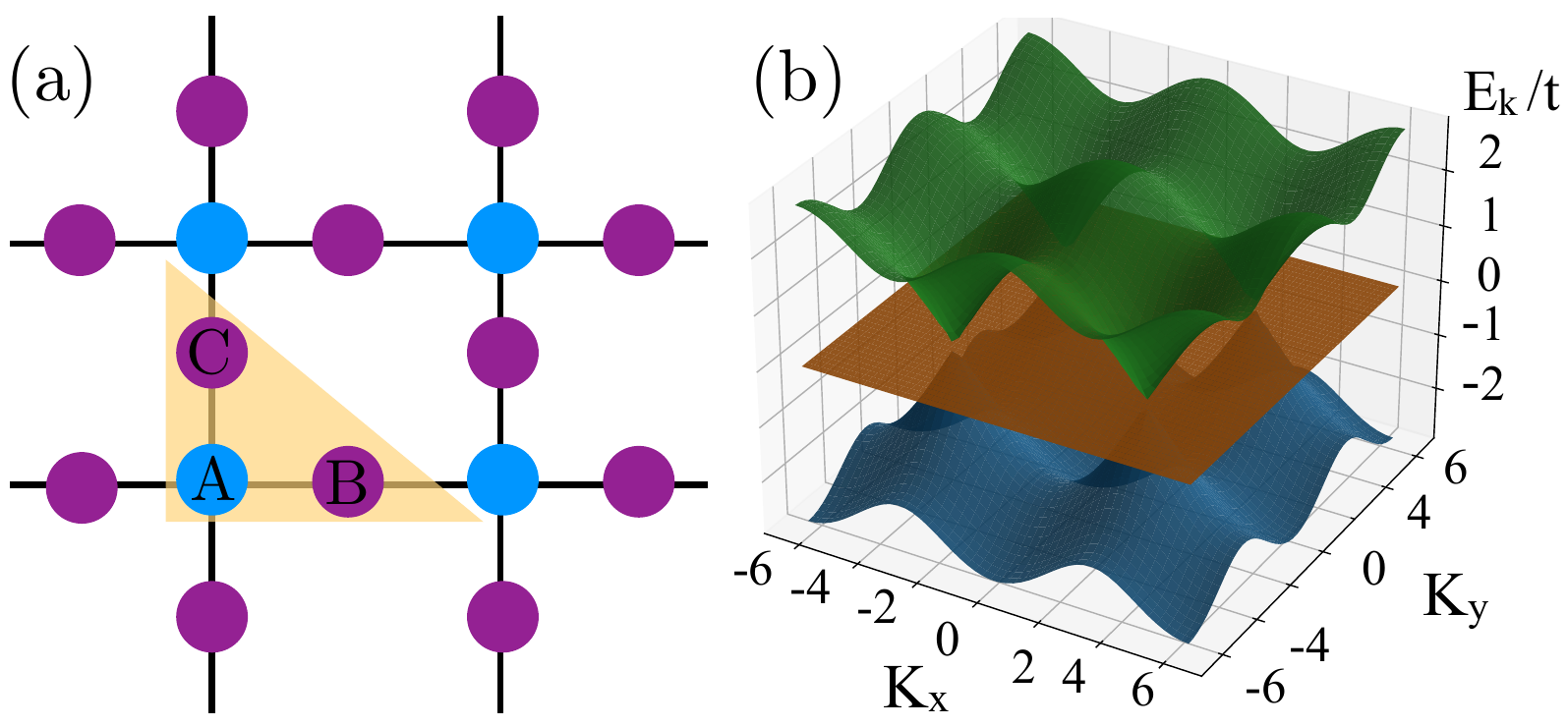}
\caption{(a) Lieb lattice structure indicating the bond and the corner sites. The shaded region indicates the three-site unit cell. (b) Non-interacting electronic dispersion of the Lieb lattice showing the dispersive and flat bands.}
\label{fig1}
\end{center}
\end{figure}

This paper concerns various facets of superconductivity (including Lifshitz transitions) in a system defined on a Lieb lattice. The Lieb lattice is a prototypical system whose non-interacting limit hosts a flat band. Superconductivity in systems hosting a flat band is currently in vogue in the study of strongly correlated quantum materials \cite{peotta_natcom2015,liang_prb2017,julku_prl2016}. Due to the quenching of the kinetic energy scale (within the flat band), even a small electronic interaction is sufficient for strong correlation physics to play out in these systems \cite{mk_rev_arxiv2025}. Superconductivity and superfluidity in the Lieb lattice have been investigated using a wide range of numerical tools, ranging from the mean field theory (MFT) to sophisticated techniques such as dynamical mean field theory (DMFT), determinant quantum Monte Carlo (DQMC), static path approximated (SPA) quantum Monte Carlo etc.\cite{peotta_natcom2015,liang_prb2017,liang2_prb2017,liang_pra2019,julku_prl2021,huhtinen_prb2022,pentilla_comphys2025,jiang_pnas2024,julku_prl2016,scalettar_prb2014,tylutki_prb2018,huhtinen_prbl2021,swain_prr2020}. In flat band systems, the key physical quantities such as the super-fluid weight, the critical temperature, and the upper-critical field are renormalised by the quantum metric and thus remains finite even in the true flat-band limit where the effective mass ($m^*$) diverges \cite{peotta_natcom2015, julku_prl2016}. In a seminal development, Chen and Law constructed the full Landau-Ginzburg-Wilson theory for flat-band superconductors \cite{chen} and found that, in flat-band systems, the role of the effective mass of the Cooper pairs $m^*$ is renormalized by the quantum metric. In the context of Lieb lattices, this result of Chen and Law generalizes the geometric superfluid weight that was introduced in Ref.~\cite{julku_prl2016}.

Here we study an attractive  Hubbard model defined on a Lieb-lattice where the itinerant electrons are coupled to classical core spins via a Kondo coupling. This is the multi-band generalization of the single-band square lattice model studied in the presence of magnetic degrees of freedom, typified by the localized classical spins that interact with the itinerant fermions\cite{mk_prb2016}. The interplay of magnetic and superconducting fluctuations in the square lattice model unravels a plethora of interesting ground states: They include phases where magnetism of the core spins coexisting with $s$-wave superconductivity, and exotic superconducting phases such as the FFLO phase \cite{fuldeferrell1964,lo1964}. 

Thus, in this paper we explore the ground states that emerge due to the interplay of magnetism and superconductivity in multi-band systems such as the Lieb lattice. More importantly, we show that the model encodes two types of Lifshitz-type transitions: In the limit where the Kondo coupling is tuned to zero, we show that the minimum energy surface undergoes a Lifshitz type transition, that is controlled by the chemical potential $\mu$. In other words, the topology of the minimum energy surface changes as a function of the chemical potential $\mu$. Furthermore, for finite Kondo coupling, we conclusively show that the dimensionality of the minimum energy surface changes, (for certain chemical potentials), as a function of the Kondo coupling. More specifically, we show that for certain conditions, as the Kondo coupling is increased, the minimum quasiparticle energy is driven to zero, leading to the emergence of nodes on the Fermi surface. \\

Even though our results are couched in terms of the physics of Lieb lattice systems, our approach is very general and can be used to study the impact 
of multi-band physics, (especially as pertained to Lifshitz like physics) in a wide variety of systems. The remaining of this paper is structured as follows: Sec.~\ref{sec:MMI} discusses the Hamiltonian which models the system, the numerical and analytic formalism used to address the same, followed by Sec.~\ref{sec:PD} containing the results and analysis of the thermodynamic properties of this system, wherein the phase diagram of the model is elucidated. The spectroscopic signatures and the discussion of a $\mu$-driven Lifshitz type transition for the case where the Kondo coupling is tuned to zero is discussed in Sec.~\ref{sec:J0}. The spectroscopic signatures for the case with finite non-zero Kondo coupling are discussed in Sec.~\ref{sec:Jneq0}. This section also elucidates the nature of the Lifshitz-type transitions that occur in such systems. We conclude with Sec.~\ref{sec:conc} that summarizes our findings. Details on the calculations for the analytic formalism are presented in the appendix.

\begin{figure}
\begin{center}
\includegraphics[height=4.5cm,width=8.5cm,angle=0]{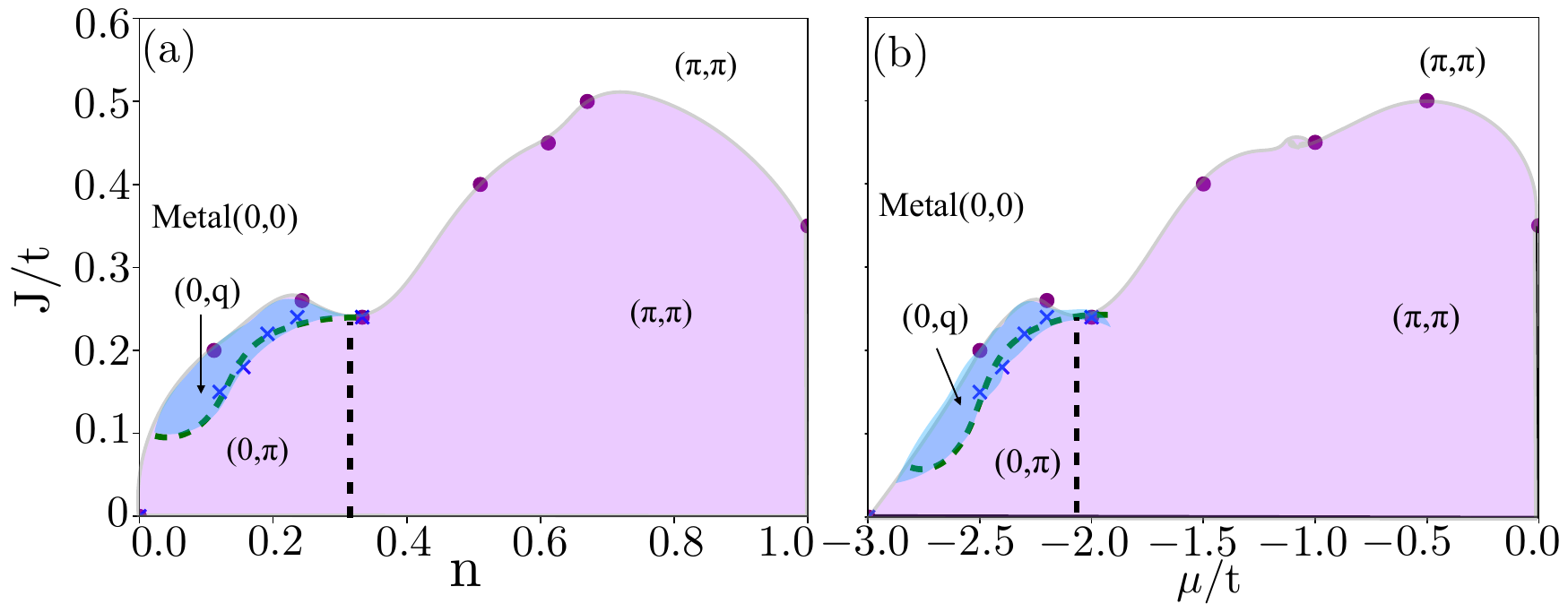}
\caption{Ground state ($T=0.01t$) phase diagram for magnetic superconductor in the (a) $n-J/t$ and (b) $\mu/t-J/t$ plane, demarcating the 
$s$-wave magnetic superconductor and magnetic metal. The phases with collinear ($\pi, \pi$), ($0, \pi$) magnetic order corresponds 
to gapped superconducting state while the superconducting regime with non-collinear ($0, q$) order is gapless.}
\label{fig2}
\end{center}
\end{figure}

\section{Model, Method and Indicators}
\label{sec:MMI}

Our starting Hamiltonian is the 2D attractive Kondo-Hubbard model on a Lieb lattice \cite{lieb_prl1989}, 
\begin{eqnarray}
H &=& -t\sum_{\langle ij \rangle,\sigma}(c_{i,\sigma}^\dagger c_{j,\sigma} + H.c)-\mu \sum_{i,\sigma} \hat{n}_{i,\sigma} - |U| \sum_{i} \hat{n}_{i,\uparrow} \hat{n}_{i,\downarrow} \nonumber \\ && - J \sum_{i} \Vec{S_{i}} . \vec{\sigma_{i}}
    \label{eq:Ham}
\end{eqnarray}
where $t=1$ is the nearest-neighbor hopping amplitude and sets the reference energy scale of the problem.  The two dimensional Lieb lattice is a bi-partite system with a three atom unit cell comprising of two bond-sites and a corner site,  \cite{lieb_prl1989}, see Fig.\ref{fig1}(a). $|U|$ encodes the attractive Hubbard interaction between the fermions, leading to an $s$-wave superconductivity. In a similar vein, the $J$ encodes the ferromagnetic Kondo term that couples the classical core-spins to the itinerant fermions hopping on the Lieb lattice\cite{Jnote}. In the non-interacting limit, ($U=J=0$), the system can be diagonalized in a band basis giving rise to flat band (having support stemming from purely the corner sites), and two dispersive bands that touch at the $M$ point as seen in Fig.\ref{fig1}(b) \cite{lieb_prl1989}.\

The results presented in this manuscript correspond to the ground state of the system, which could be suitably accessed via a real-space Bogoliubov-de Gennes (BdG) Mean Field Theory (MFT) by performing a Hubbard-Stratonovich (HS) transformation \cite{hubbard,stratonovich}. The attractive on-site Hubbard interaction in terms of the superconducting pairing field $\Delta_i$. Therefore, Eq.~\ref{eq:Ham} can be written as:
\begin{eqnarray}
H_{eff} &=& -t\sum_{\langle ij \rangle,\sigma}(c_{i,\sigma}^\dagger c_{j,\sigma} + H.c)-\mu \sum_{i,\sigma} \hat{n}_{i,\sigma}+ J \sum_{i} \Vec{S_{i}} . \Vec\sigma_{i}  \nonumber \\ && + \sum_{i} (\Delta_{i} c_{i\uparrow}^\dagger c_{i\downarrow}^\dagger+\Delta_{i}^* c_{i\downarrow} c_{i\uparrow}) + \frac{\vert\Delta_{i}\vert^2}{\vert U\vert} 
\label{eq:Heff}
\end{eqnarray}
In this work, the superconducting pairing field that couple to the fermions considered purely real and treated as a site independent quantity with in the mean field analysis. The magnetic order is treated by introducing spiral state configurations such that, the polar angle is $\theta_{i}=\pi/2$ 
and the azimuthal angle $\phi_{i}=({\bf q.}{\bf r}_{i})$ is periodic as, $S_{i}^{z}=0$, $S_{i}^{x}=\cos({\bf q.}{\bf r}_{i})$ and $S_{i}^{y}=\sin({\bf q.}{\bf r}_{i})$, where, 
${\bf q}$ is the magnetic wave vector. Furthermore, from here on we assume that the magnitude of the classical core spins is held fixed at $S=1$.
Operationally, a MF variational calculation is performed by sitting at a low temperature, (here $T=0.01t$), and optimizing the free energy over a restricted family of $\{{\bf S}_{i}, \Delta_{i}\}$ configurations. The magnetic wave vector is defined as, $\{q_{x}, q_{y}\} = 2\pi i/L$ where, $i=0, 1, 2, ...$. The free energy minimization is carried out over $\{q_{x}, q_{y}\}$ and $\vert \Delta\vert$ for the $\mu- J$-space and at the fixed pairing interaction of $\vert U\vert=2t$, unless specified otherwise. The calculations are carried out in the grand canonical ensemble, and the $\mu$ values are converted to the corresponding lattice filling ($n$) so as to present the results in the $n-J$-plane. The typical lattice size used for our calculations is $L=20$\cite{qsoftnote}. 

\begin{figure}
\begin{center}
\includegraphics[height=7.5cm,width=8.4cm,angle=0]{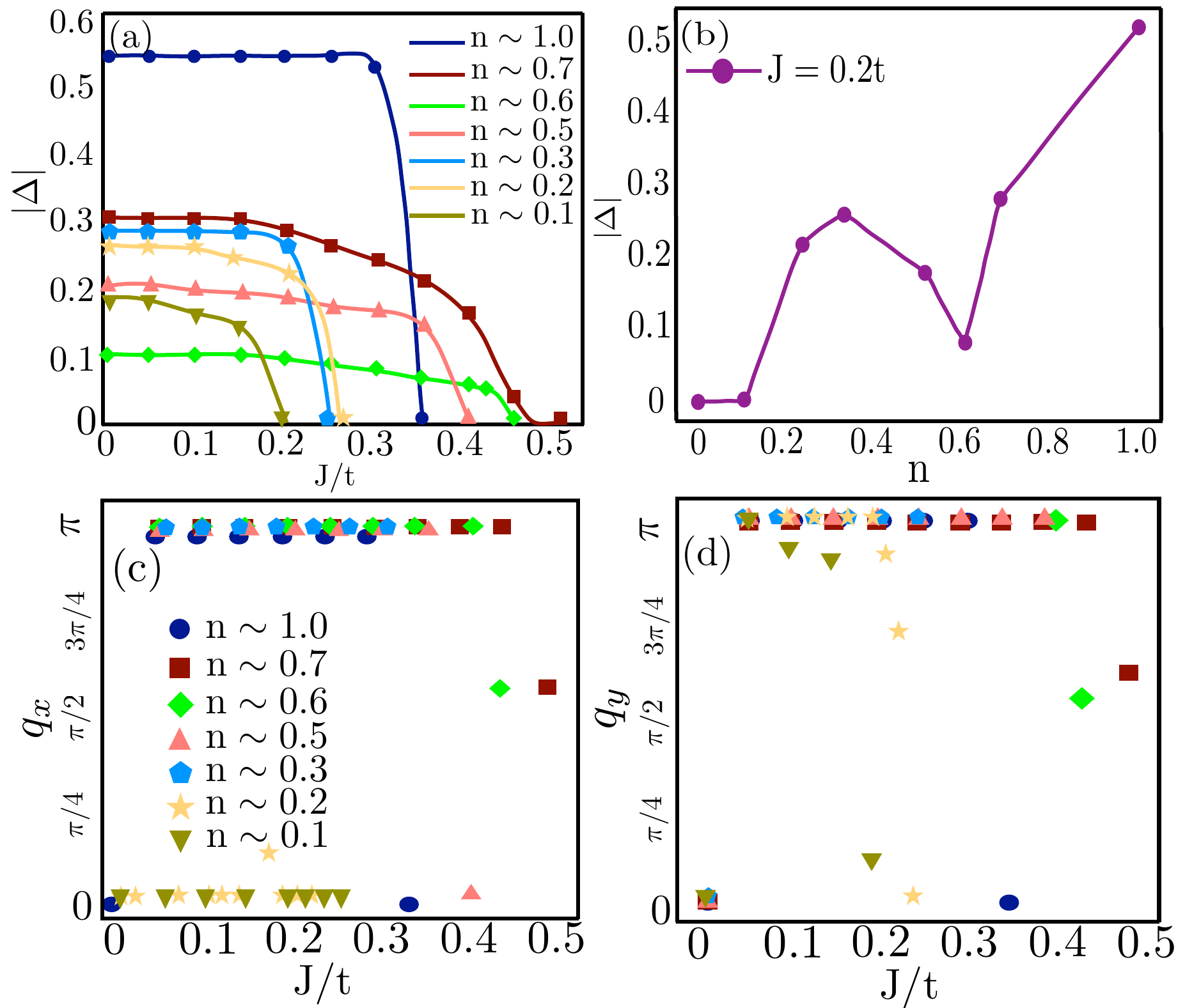}
\caption{Thermodynamic observables quantifying the magnetic superconductor. (a) Variation of the superconducting pairing field 
amplitude with the Kondo interaction at selected fermion density. (b) Non-monotonic variation of the superconducting pairing field amplitude 
w. r. t. $n$, shown at a representative Kondo interaction of $J=0.2t$. (c)-(d) Variation of the components of the magnetic 
wave vector with the Kondo interaction at selected $n$. Panels (c) and (d) share the legend.}
\label{fig3}
\end{center}
\end{figure}

In addition to the mean-field variational approach, the spectroscopic signatures of this system are analyzed using a momentum-space Green's 
function formalism, which provides an analytic handle to access the quasiparticle scattering at the mean-field level beyond the finite-size restrictions and supports the numerical in the thermodynamic limit. The technical details concerning the development of spectroscopic properties discussed in this work, such as constructing the eigenvalue spectrum and the role of inter/intra band contributions in momentum space, are presented in the Appendix.
\begin{figure}
\begin{center}
\includegraphics[height=3.5cm,width=8.2cm,angle=0]{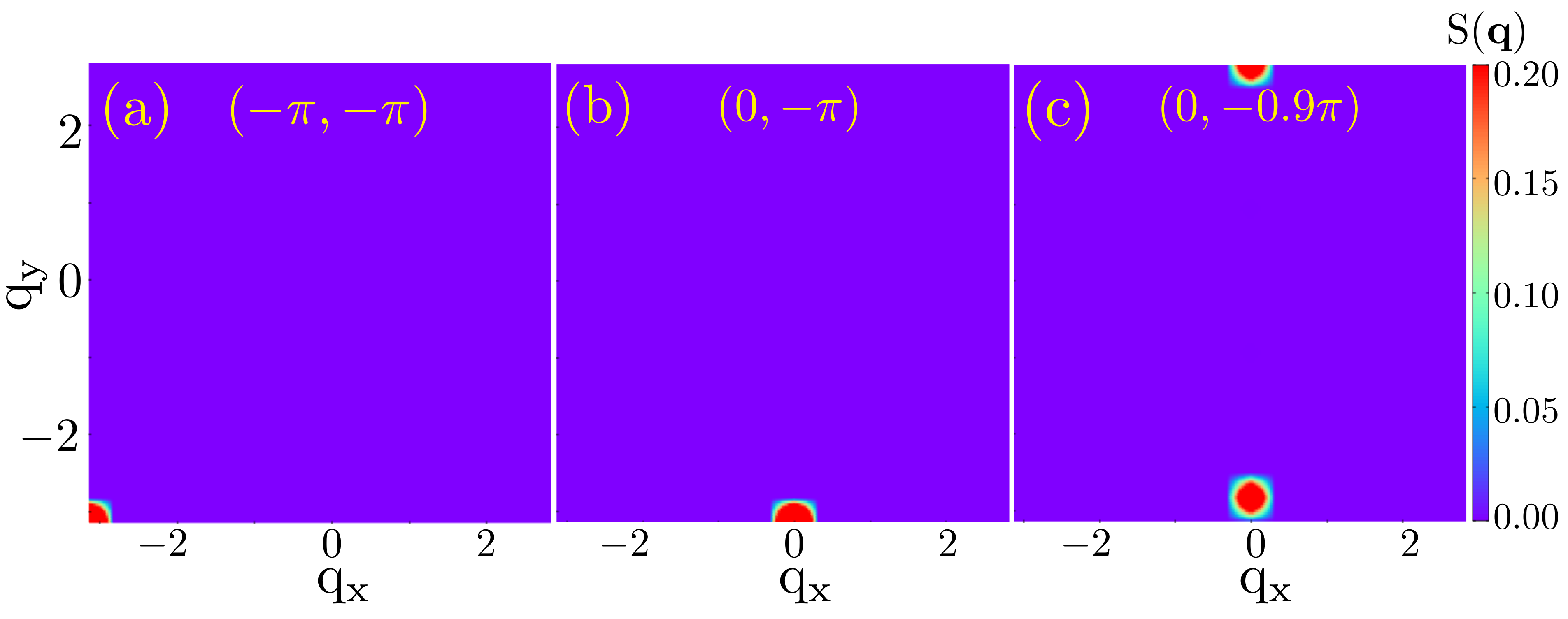}
\caption{Magnetic structure factor maps highlighting the magnetic ordering at the selected $n-J$ cross sections of (a) $n \sim 0.6, J=0.1t$, 
(b) $n \sim 0.24, J=0.1t$ and (c) $n \sim 0.3, J=0.24t$.}
\label{fig4}
\end{center}
\end{figure}
 
The thermodynamic phases and transitions in this system are quantified based on the: (i) magnetic structure factor $S({\bf q})=\frac{1}{N}\sum_{ij}\langle{\bf S}_{i}.{\bf S}_{j}\rangle e^{i{\bf q.r}}$, where $N=3L^{2}$ is the number of lattice sites, (ii) momentum-resolved spectral function 
\begin{eqnarray}\label{eqn:spectalfun}
   A({\bf k}, \omega) = -\frac{1}{\pi}{\mathrm {Im}} [ G({\bf k}, \omega)], \nonumber \\ 
   G({\bf k}, \omega) = \lim_{\delta \rightarrow 0} G({\bf k}, i\omega_{n})\vert_{i\omega_{n}\rightarrow \omega + i\delta}, 
\end{eqnarray}
    
where $G({\bf k}, i\omega_{n})$ is the imaginary frequency transform of $\langle c_{k}(\tau)c^{\dagger}_{k}(0)\rangle$,  (iii) single-particle density of states (DOS) $N(\omega)=\sum_{\bf k}A({\bf k}, \omega)$\cite{unweighted}, and (iv) energy gap $E_{g}$,  as obtained from the minimum eigenvalues ($E_{n}$) of the BdG spectrum.

\section{Phase diagram and thermodynamic signatures}
\label{sec:PD}
We begin the discussion with the ground state phase diagram in the $n-J$-plane, Fig.\ref{fig2}(a), which sums up the primary thermodynamic results of this work. The same has been depicted in the $\mu-J$-plane in Fig.\ref{fig2}(b).  For the remaining of this paper we discuss the results in terms of $n$ or $\mu$, alternatively.  In the absence of Kondo interaction ($J=0$) the system hosts a gapped $s$-wave superconducting state across the lattice filling, quantified in terms of a finite superconducting pairing field amplitude $\vert \Delta\vert \neq 0$ and a robust single particle spectral gap ($E_{g} \neq 0$) at the Fermi level. A {\it magnetic superconductor} is realized for $J/t \neq 0$ such that,  the superconducting order coexists with (quasi) long range magnetic correlations. 
Over a large part of the phase diagram, $0.3 < n \le 1.0$ ($-2t \lesssim \mu < 0 $) $s$-wave superconductivity survives with a collinear $(\pi, \pi)$ magnetic order.
 which gives way to a collinear $(0, \pi)$ order over the regime $0 < n \le 0.3$ ($-3t < \mu \lesssim -2t$). Stronger Kondo coupling ($0.1t \lesssim J < 0.25t$) leads to gapless $s$-wave superconductivity, as observed for $0 < n \le 0.3$. This gapless magnetic superconductor is characterized by a finite superconducting pairing field amplitude, $\vert \Delta\vert \neq 0$ and a spectral gap closure ($E_{g} = 0$), along with non-collinear magnetic order with magnetic wave vector $(0, q)$. Further increase in Kondo interaction gives way to a magnetic metal with finite magnetic wave vector and an absence of superconducting correlations. Later in this section we will elucidate the reasoning for why only the $(0,q)$ ordered state is associated with gapless superconductivity.

\begin{figure*}
\begin{center}
\includegraphics[height=3.8cm,width=15.5cm,angle=0]{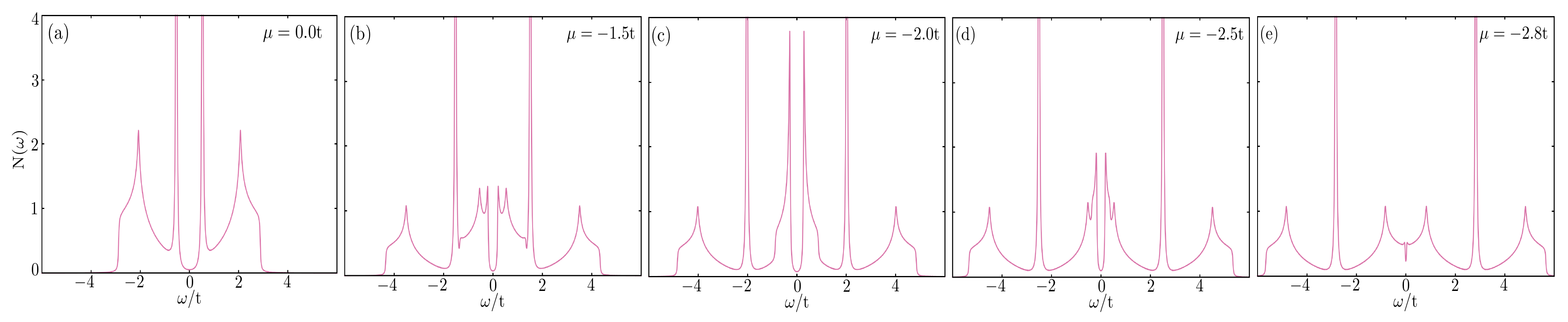}
\caption{(a)-(e) Single particle DOS ($N(\omega)$) as obtained via the Green's function formalism at $J=0$ and different chemical potential 
$\mu/t$. }
\label{fig5_0}
\end{center}
\end{figure*}
\begin{figure*}[!t]
\centering

\includegraphics[width=15.5cm]{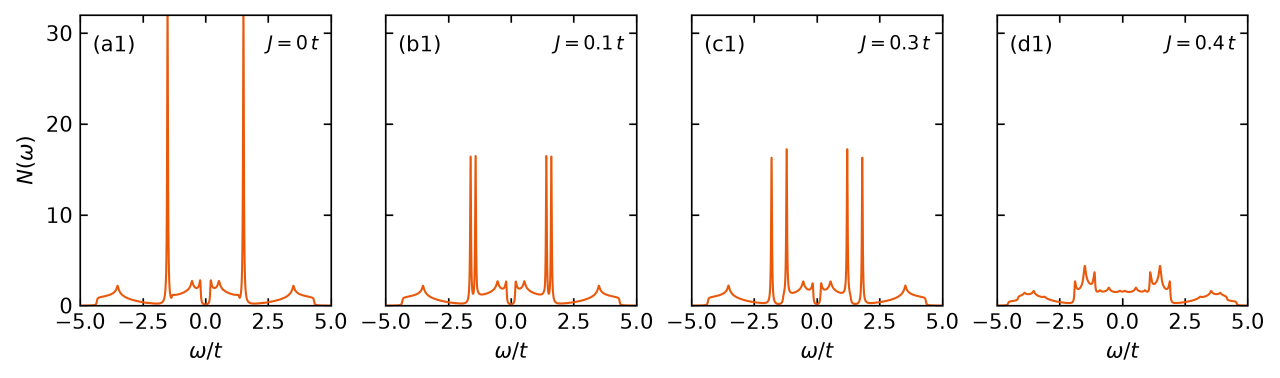}\\[-1mm]
\includegraphics[width=15.5cm]{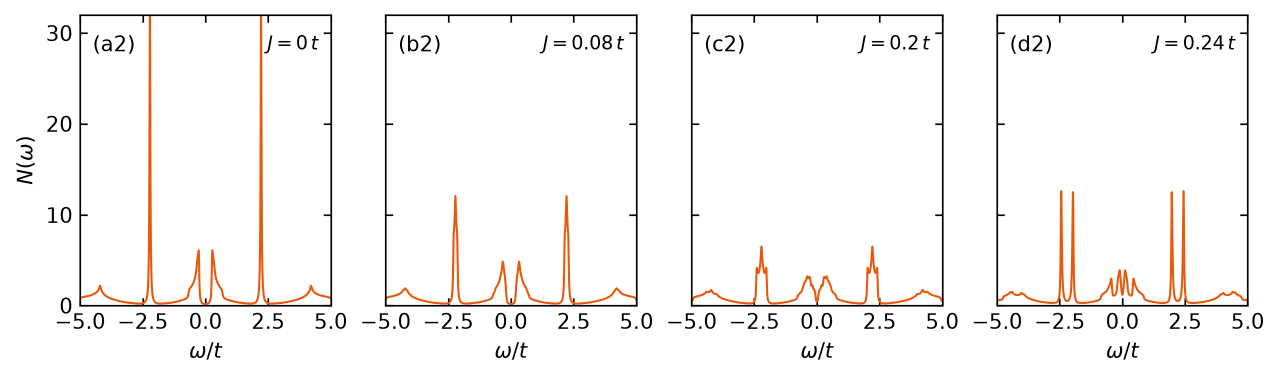}

\caption{Single particle DOS at selected $J/t$ showing the impact of the
Kondo interaction at $\mu=-1.5t$ and $\mu=-2.2t$. As elucidated in Sec.~\ref{subsec:Dos_Jneq0}, for $\mu=-1.5t$, the DoS ((see panels (a1)-(d1))clearly shows the effect of the predominant $(\pi,\pi)$ order that obtains here for finite $J$ wherein the system is always gapped and the flat-bands effectively gives Zeeman split $\delta$-functions with the splitting exactly equal to $2J$.At $J=0.4t$, the spectral gap has collapsed and the system is a magnetic metal supporting $(0,\pi)$ order. In contrast, the DoS at finite $J$ for the $\mu=-2.2t$ (panels (a2)-(d2) clearly shows the spectral gap collapsing, with increasing $J$. This closing of the spectral gap is pinned to incommensurate ordered phases. See Sec.~\ref{subsec:Dos_Jneq0}, for further details.}
\label{dosjneq0}

\end{figure*}

The thermodynamic signatures quantifying these phases are presented in Fig.\ref{fig3} and Fig.\ref{fig4}. The variation 
of the superconducting pairing field amplitude $\vert \Delta \vert$ with $J/t$ at selected $n$ are shown in Fig.\ref{fig3}(a). At the half filling 
($n=1$) a sharp first order phase transition takes place between a magnetic metal and a magnetic superconductor with the magnetic wave vector being $(\pi, \pi)$,  across the critical Kondo coupling $J_{c1} \sim 0.35t$. Away from the half filling the suppression of $\vert \Delta\vert$ with $J/t$ is rather continuous, indicating a second order phase transition between the magnetically correlated metal and the superconducting phases. The underlying multiband lattice structure however, leads to a non monotonic dependence of $\vert \Delta\vert$ on $n$,  as is evident from Fig.\ref{fig3}(b). This is in sharp contrast with the observations made on the single band Kondo Hubbard model on the square lattice with monotonic dependence of the superconducting pairing field amplitude on the lattice filling \cite{mk_prb2016}. We note that the observed non-monotonicity 
of the superconducting order is purely due to the multiband structure of the Lieb lattice and is observed at $J=0$ as well, in agreement with the 
inferences drawn based on DMFT \cite{julku_prl2016} and DQMC \cite{scalettar_prb2014} simulations. 

At selected lattice filling the variation of the magnetic wave vector components with the Kondo interaction $J/t$ is shown in Fig.\ref{fig3}(c) and Fig.\ref{fig3}(d). Over the regime $0.3 \lesssim n < 1$, $q_{x}$ and $q_{y}$ behaves identically, giving rise to a $(\pi, \pi)$ magnetic order while for $0 < n \lesssim 0.3$;  a $(0, \pi)$ and a $(0, q)$ magnetic order are realized at weak and strong Kondo interactions, respectively. Fig.\ref{fig4} presents the 
magnetic structure factor ($S({\bf q})$) maps at the representative $n-J$ cross sections, highlighting the variation in the magnetic wave vector.  
A $(-\pi, -\pi)$ peak in the $S({\bf q})$ map at $n \sim 0.6$, $J=0.1t$ (Fig.\ref{fig4}(a)) corresponds to the underlying collinear magnetic 
order while at $n \sim 0.24$ and $J=0.1t$ (Fig.\ref{fig4}(b)) a $(0, -\pi)$ magnetic order is stabilized. A non-collinear ($0, -0.9\pi$)\cite{kgrid} magnetic order 
is observed at $n \sim 0.3$ and $J=0.24t$ (Fig.\ref{fig4}(c)). 

In what follows, we will investigate the existence of gapless phases for various ordering wave vectors that we have seen above. 
Our explanation hinges on the realization that the existence of the gapless phase is tied to the determinant of the Hamiltonian written in the momentum 
space formulation of the MF BdG description (defined as ${\cal H}$ in Eq.~\ref{eq:H11}).  Specifically, Eq.~\ref{eq:detsplit} states that the the determinant of ${\cal H}$ can be 
written as a sum of squares of two real polynomials $R$ and $I$. Thus, the requirement for the gapless phase implies that both $R$ and $I$ should be zero, for a range of values of the momenta 
or at isolated points in the Brillouin zone such that a gapless contour or isolated nodes obtains.

First we turn our attention to the question of why the gapless state cannot be associated with the commensurate $(0,0)$ and $(\pi, \pi)$ ordered state.  From the discussion in the Appendix (see Sec.~\ref{app:ferro}, and Table~\ref{tab:bzsummary-long}), the spectrum allows for a ferromagnetic solution $q\equiv(0,0)$, which could become gapless when the effective  Kondo coupling $h=J|S|/2$, attains the critical $J_c=\Delta$, (for the case $S=1$, see Appendix for further details). This can be easily understood if one realizes that the system with the ferromagnetically ordered core spins maps onto the case of an s-wave superconductor in a uniform Zeeman field. However, this co-existent singlet superconductor with the ferromagnetically ordered state is a maximum of free energy, and thus we do not observe it here in the phase diagram\cite{sarma1963,liuwilczek2003}.

We now argue that the $(\pi,\pi)$ ferrimagnetic order is generically always gapped. The ferrimagnetic case is entry II in Table~\ref{tab:bzsummary-long},
which clearly states that there are no generic dispersive nodes at any momentum, and the whole Brillouin zone could go gapless as far as the flat-bands are concerned. 
Below, we explain the rationale behind this entry in Table~\ref{tab:bzsummary-long}: For the ferrimagnetic ordering wave vector, it is clear that $I\equiv 0$, (see \ref{subsec:gapclpipi}). However, from Eq.~\ref{eq:detpipi}, $R=0$ is only possible if the quantity 
$\Lambda$ of Eq.~\ref{eq:Lambda}, vanishes. This condition $\Lambda =0$ gives a momentum-independent condition that clearly states that the whole Brillouin zone becomes gapless in the flat-band sector if the condition $h^{2}=\mu^{2}+\vert\Delta\vert^{2}$ is met. Another restatement of the above condition for obtaining gapless states in the flat-band sector is to 
couch it in the language of the spectral gap: The spectral gap in the flat band sector is given $\left|\sqrt{\mu^2+|\Delta|^2}-h\right|$, implying, as we have already seen that the entire Brillouin gapless at a single point, and past the point the gap reopens. Since this gaplessness condition is extremely fine tuned we state that this is as a non-generic case. Finally, Eq.~\ref{eq:detpipi} also clearly states that the dispersive bands generically do not provide a gapless phase. However, at the isolated $M$ point where $E_{\bf k}=0$ is satisfied one expects a dispersive node. 

The $(\pi,\pi)$ state therefore remains generically a fully gapped $s$-wave magnetic
superconductor throughout its region of the phase diagram. Nowhere is this phenomenology of the fully gapped nature at the $(\pi,\pi)$ order more evident than the case $\mu=0$,
(corresponding to the case of the half-filled lattice). Here, the flat bands together give a contribution to the free energy which is exactly $-\Delta$, (the dispersive band provide an $O(J^2)$ contribution
to the free energy). In a similar vein, in the non-superconducting part ($\Delta=0$) of the phase diagram, in the half-filled limit, the flat bands provide a $-J$ contribution to the free energy. Thus, one has a mechanism where at $J=\Delta$, two straight-lines with different slopes cross providing a mechanism for a first order transition that is seen in Fig.~\ref{fig3}.

We now look at the stripe $(0,\pi)$ state (entry III of Table.~\ref{tab:bzsummary-long}). The entries corresponding to this wave clearly highlight that there are three possible classes of zero modes that are possible for the stripe-ordered state. This is elucidated in great detail in Sec.~\ref{app:0pi} of the Appendix. However, a complete and 
detailed investigation of the same here, including a re-elucidation of the gap closing condition, would be a purely academic exercise, as these gapless phases with $(0,\pi)$are not thermodynamically stable. These conditions in any case are elaborated in Sec.~\ref{app:0pi} of the Appendix.
 
Now, for the case of the $(0,q)$ order, on the locus on ${\cal L}_1$, Eq.~\ref{eq:L1} gives the condition for the existence of the nodes. Thus, as discussed in Sec.~\ref{subsec:gapclosing0q}, this would immediately imply that at most there could be four nodal points that will populate the Brillouin zones, always coming in symmetry-related pairs $(\pm k_{x}^{\star},-q)$, (two nodes for each branch of Eq.~\ref{eq:L1}). The nodal points come in pairs, and the number of such gap-closing points that finally accrue on ${\cal L}_1$ depends on whether the branches ($\pm$) lie in the interval $[0,1]$.
Furthermore, as discussed in the same section \ref{subsec:gapclosing0q}), one could have at most four nodal points that stem from ${\cal L}_2$.  Thus, for example, for the case when 
$\mu=-2.2t$, and $J=0.24t$, one can clearly show that only two nodal points exist, as the other pair correspond to a branch that lie outside the interval $[0,1]$ (see Sec.~\ref{subsec:spec_Jneq0}). 
For the case that entails the locus ${\cal L}_2$, please see Ref.~\cite{locus2}.

\section{Lifshitz like transition for $J=0$}
\label{sec:J0}

\begin{figure*}[t]
\centering
\includegraphics[width=\textwidth]{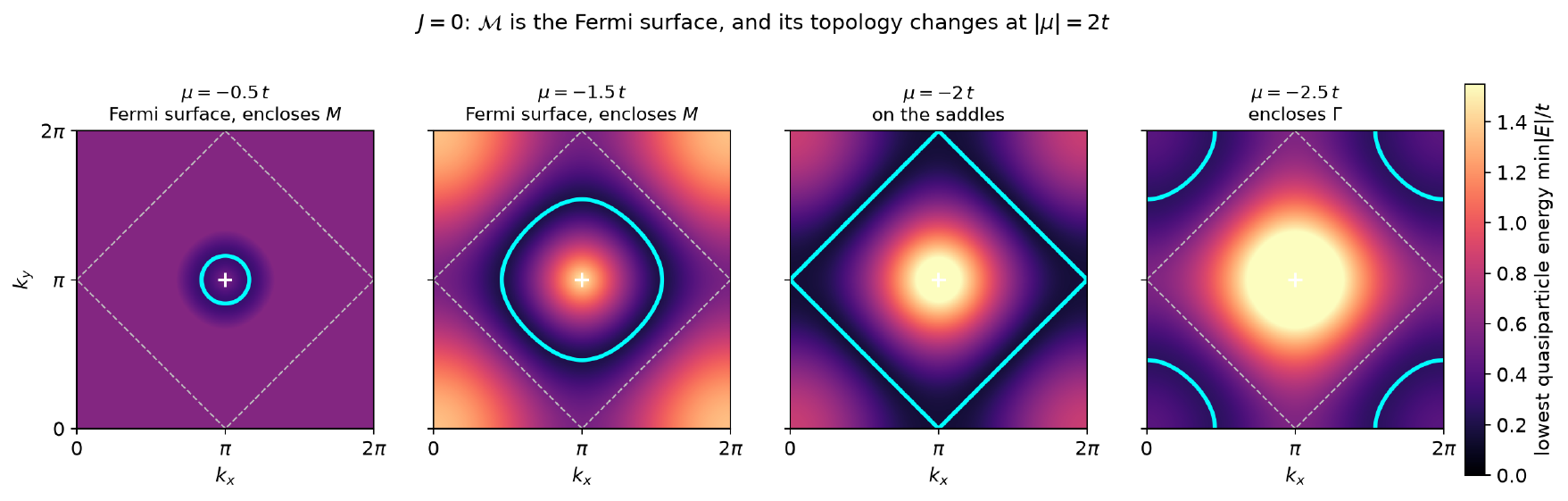}
\caption{Locus of gap minima, $\cos k_x+\cos k_y=\mu^2/2t^2-2$, on the zone
$[0,2\pi]^2$ with $M$ at the centre, $X$ at the edge midpoints and $\Gamma$ at
the corners; shading marks the occupied region. (a) $-2t<\mu<0$: a hole pocket
about $M$, drawn for $\mu=-1.5t$. (b) $\mu=-2t$: the contour is the square
$|k_x-\pi|+|k_y-\pi|=\pi$, touching the zone boundary at $X$ and perfectly
nested by $\bm{Q}=(\pi,\pi)$. (c) $-2\sqrt2\,t<\mu<-2t$: the contour has
reconnected into pockets about $\Gamma$, drawn for $\mu=-2.4t$.}
\label{fig:Lifshitz}
\end{figure*}

%
%
%

\newcommand{\lieblabel}[2]{%
  \node[font=\scriptsize,below,align=center] at (1.15,-0.12)
       {#1\\[-0.3ex]{\tiny #2}};}


We now turn our attention to the spectroscopic signatures that obtain at $J=0$ as a function of the chemical potential $\mu$.  Fig.\ref{fig5_0} shows the $N(\omega)$ and the corresponding for particular values of $\mu$. The BdG spectrum conforms to the functional form 
 $\omega = \pm ({(\epsilon_{k}-\mu)}^{2}+\Delta^{2})^{1/2}$, where the $\epsilon_k$ are the electronic dispersions that stems from the non-interacting Lieb lattice, (see Appendix). 
 The above dispersion relation clearly attests to the fact that the minimum energy condition for any finite $\Delta$, would correspond to the case where the quantity $\epsilon_{k}-\mu =0$, giving a $\mu$ dependent condition,
  \begin{equation}
  \cos k_x+\cos k_y=\frac{\mu^2}{2t^2}-2, 
  \label{eq:min_condn}
  \end{equation}
 that the minimum energy contour $\mathcal{M}$ has to satisfy. The topology of this minimum energy contour as the chemical potential $\mu$ is tuned describes the Lifshitz like transition in this multiband system: At $\mu=0$, the minimum energy contour encloses one single point at  $M\equiv(\pi,\pi)$. As seen in Fig.~\ref{fig:Lifshitz}, for $-2t<\mu<0$, the minimum energy contour is a closed curved enclosing the $M$ point. As the chemical potential hits $\mu=-2t$, the character of the minimum energy surface changes to a square $|k_x-\pi|+|k_y-\pi|=\pi$. Furthermore, at this chemical potential this square minimum energy contour is perfectly spanned by the nesting vector $Q=(\pi,\pi)$. This perfect nesting reflects itself into the phase diagram in the $\mu-t$ plane as seen in Fig.~\ref{fig2}. More specifically, the perfect nesting at $\mu=-2t$leads to a logarithmic enhancement of the Lindhard function, thus ensuring that the onset of the of the $(\pi,\pi)$ order falls at that particular value of the chemical potential. In the regime $-2\sqrt2 t<\mu<-2t$  the minimum energy contour now encloses the high symmetry $\Gamma$ point.

 The signatures of the stationary points (at $k\equiv(0,0), (0,\pi), (\pi, 0)$, and at ($\pi,\pi)$), can be seen in the behavior of the DoS earmarked in Fig.~\ref{fig5_0}: Specifically, these momenta gives us critical energies,$\epsilon_c=0, 2t$, and at $2\sqrt{2}$, which shape the structure of the DoS. For instance, for $\mu=0$, see Fig.~\ref{fig5_0}(a), the 
 he dominant peak is given by the flat bands that contribute a $\delta$ function singularity at $\omega=\Delta$. Furthermore, the conical point at $M$ provides a step singularity which is overwhelmed by the delta-function flat band contribution. Finally, the critical energy at $-2\sqrt{2}$ that stems from the band minimum contributes to the step discontinuity at the spectral edge. 
 
 Now, for any finite $\mu <0$, typified by the case $\mu=-1.5$, shown in Fig.~\ref{fig5_0}(b), the contour encircling the $M$ point provide the typical BCS square root divergence. The critical energy at $-2t$, associated with the saddle at $X$ provides the logarithmic divergence. The contribution to the DoS that from the band edge at the critical energy $\epsilon_c = -2\sqrt{2}t$, provides the step singularity whereas the delta function contribution once again arises out of the flat-band contribution. Sweeping $\mu$ to the critical value of $-2t$, that earmarks the Lifshitz transition, the DoS, one immediately sees the DoS reflects the neck collapse that is associated with the Lifshitz transition. The DoS at $\mu=-2t$, Fig.~\ref{fig5_0}(c) clearly shows the gap edge divergence 
 stemming from the closed contour enhanced with a logarithm that stems from the saddle points that live on the contour. Now, at $\mu=-2.5t$, (see Fig.~\ref{fig5_0}{(d)}), the gap edge singularity is of the usual BCS form. Furthermore, the minimum at $\Gamma$, associated with $\epsilon=-2\sqrt{2}t$, provides the step-down discontinuity and the saddle at $X$ provides the $\ln$ divergence. Finally, the $\delta$ function effectively stems from the flat-band contribution. Last, we turn our attention to the case $\mu=-2\sqrt{2}$, which lies on the edge of the non-interacting Lieb lattice tight-binding model. The critical energy $\epsilon_c=-2\sqrt{2}$ with the minimum at $\Gamma$ gives a step contribution to the DoS. Furthermore, the saddle at $X$ gives the log divergence, and the flat bands at $\epsilon_c=0$ gives the overwhelming $ \delta$-function signatures. The signature of the Lifshitz transition is clearly seen in Fig.~\ref{fig5_0}(c), where the square root divergence at the band edge is renormalized with a logarithmic divergence (cut off by the superconducting OP).

  \begin{figure*}[!t]
\centering
\includegraphics[width=15.5cm]{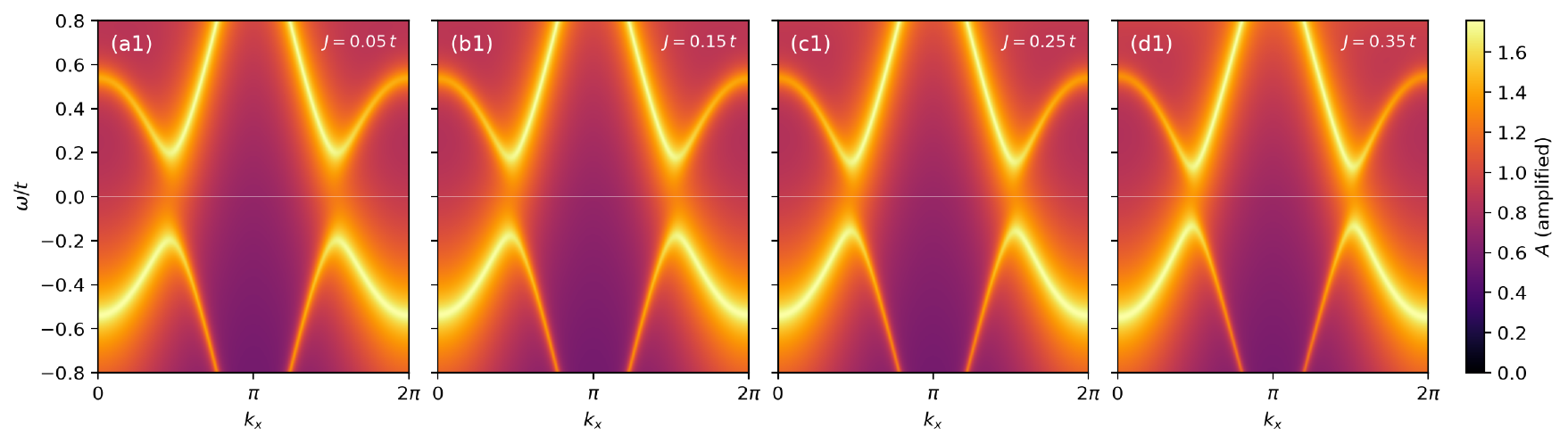}\\[-1mm]
\includegraphics[width=15.5cm]{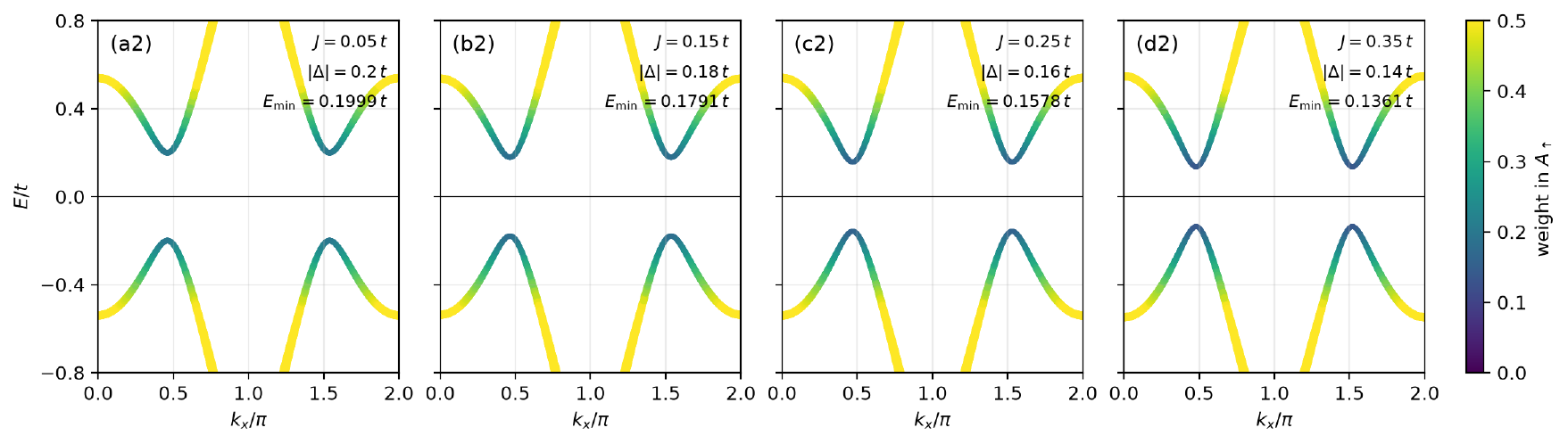}
\caption{The spectral function $A_{\uparrow}(k_x,\omega)$ at $\mu=-1.5t$, for a sequence of Kondo coupling with increasing $J/t=0.05, 0.15,0.25$, and $0.35$, (panels (a1)-(d1). In particular, the figure shows the spectral function that is plotted for $k_y=\pi$, and the $|\omega | \le 0.8t$ range is so-chosen to highlight the lowest particle hole pair that falls within that window. The graphs in conjunction with Fig.~\ref{fig5}, (panels (a2)-(d2)), which plots the exact eigenvalues of the BdG spectrum that fall within the $\omega$ window  $\mu=-1.5t$, one sees that as the coupling $J$ is tuned the contour of minimum energy is pinned to $\Mset=1$. See text for further details.}
\label{fig5}
\end{figure*}

 \begin{figure*}[!t]
\centering
\includegraphics[width=15.5cm]{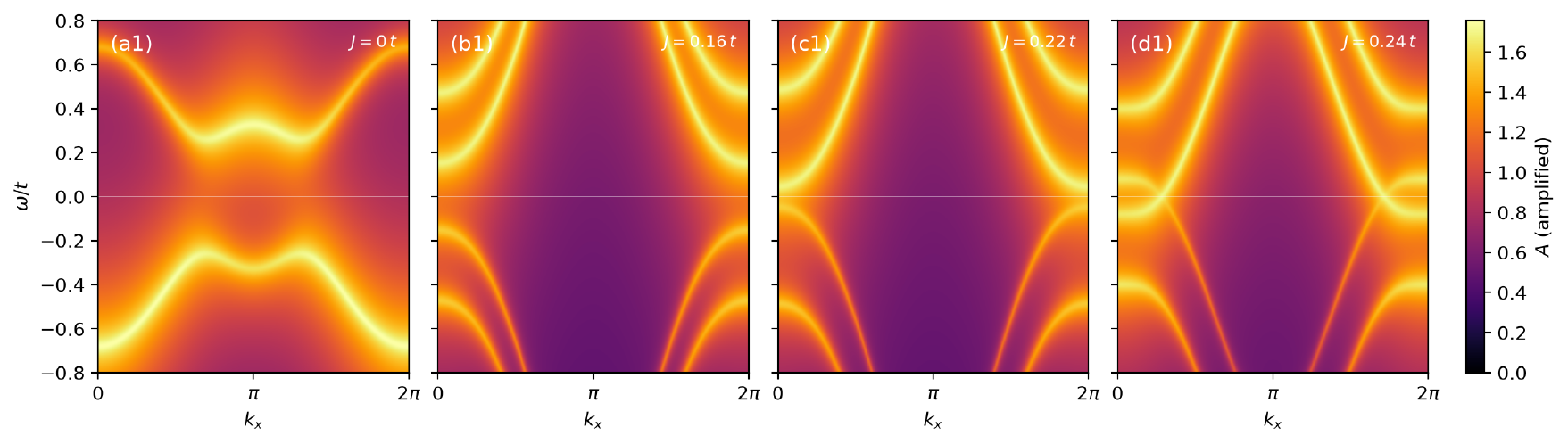}\\[-1mm]
\includegraphics[width=15.5cm]{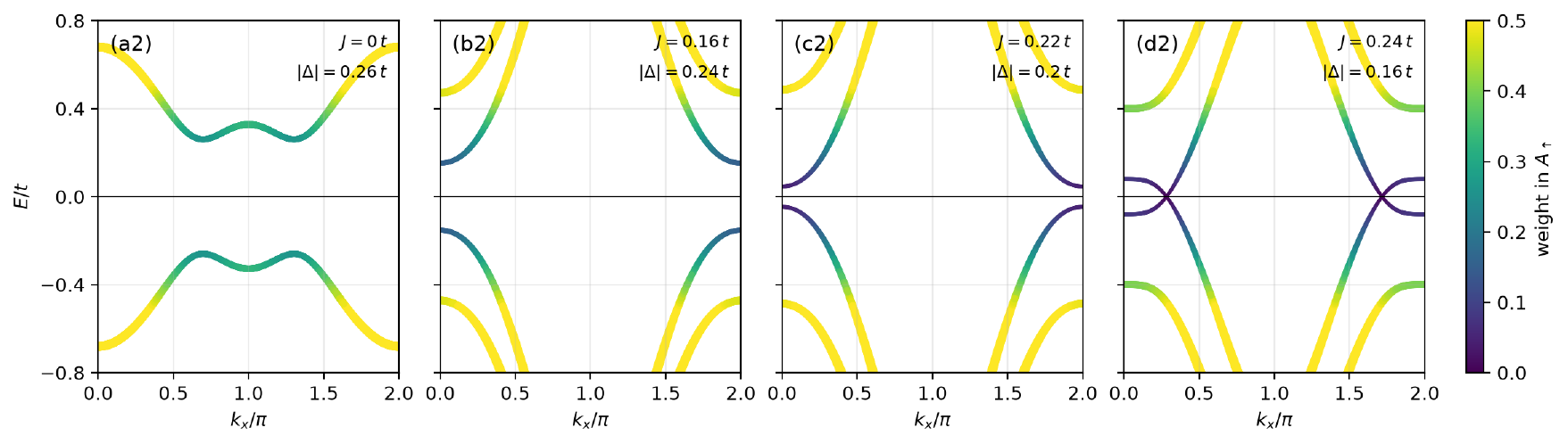}
\caption{The spectral function $A_{\uparrow}$ at $\mu=-2.2t$ is shown in panels (a1)-(d1). Once again for the sake of clarity, we use window $|\omega | \le 0.8t$, and fix $k_y=q$, (see text for further details).  In contrast to Fig.~\ref{fig5} above, with increasing $J$, one sees that a contour of minimum energy points morphs into a nodal minimum. The above is well matched with Fig.~\ref{fig:spec2.2}, (panels (a2)-(d2)) which clearly shows the emergence of a nodal state in the lowest particle-hole eigenvalue pair as the Kondo coupling is tuned, (see text for further details).}
\label{fig:spec2.2}
\end{figure*}

\section{Lifshitz transition for $J \neq 0$:} 
\label{sec:Jneq0}
\subsection{The $J$ driven transition}
\label{subsec:spec_Jneq0}
\begin{figure*}[!t]
\centering
\includegraphics[width=15.5cm]{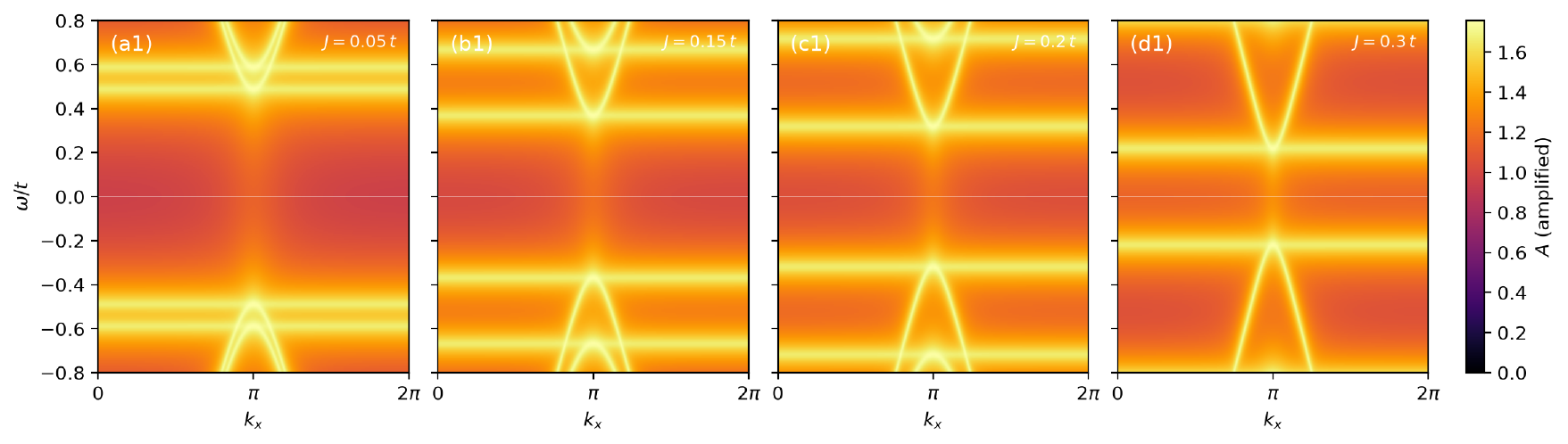}\\[-1mm]
\includegraphics[width=15.5cm]{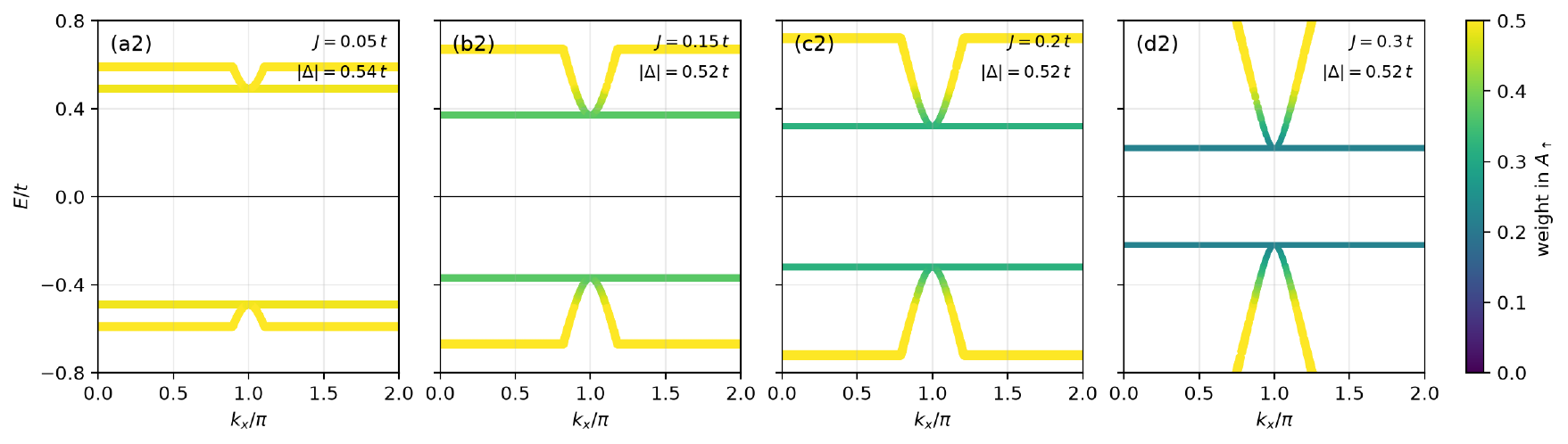}
\caption{The spectral function $A_{\uparrow}$ at $\mu=0$ is shown in panels (a1)-(d1). Once again for the sake of clarity, we use window $|\omega | \le 0.8t$, and fix $k_y=q$, (see text for further details).  At this chemical potential the flat bands come into play, and the minimum energy is given by $\vert\Delta-h\vert$. The tuning of $J$ will not give rise to the formation of nodal points.
The panels (a2)-(d2)) clearly is a retelling of the same fact but this time through the lens of the eigenvalues of the BdG spectrum, (see text for further details).}
\label{fig7}
\end{figure*}

Just like in the case of the $s$-wave superconductor defined on a square lattice \cite{mk_prb2016}, the Kondo interaction leads to two significant changes in the single particle DOS: (a) the superconducting gap at the Fermi level is progressively suppressed with increasing $J/t$ and (b) the singularities at/close to the gap edges and away from $\omega=0$ undergoes magnetic interaction induced splitting. We understand these observations as follows: for a conventional $s$-wave pairing, the quasiparticle scattering takes place between $\vert {\bf k}_{\uparrow}\rangle$ and $\vert {\bf -k}_{\downarrow} \rangle$ states and opens up the superconducting gap at the Fermi level. On the other hand, in the presence of Kondo coupling, the quasiparticles undergo finite-momentum scattering and connects a larger set of states such that, in addition to $\vert {\bf -k}_{\downarrow} \rangle$, the $\vert {\bf k}_{\uparrow}\rangle$ state now undergoes scattering from the $\vert {{\bf{-k+2q}}}_{\downarrow}\rangle$ and $\vert {{\bf{-k-2q}}}_{\downarrow}\rangle$ as well. This is in contrast to the case of case of the equivalent problem defined on the square lattice where $\vert {\bf -k}_{\downarrow} \rangle$, the $\vert {\bf k}_{\uparrow}\rangle$ states now undergoes scattering from $\vert {{\bf{-k+q}}}_{\downarrow}\rangle$ and $\vert {{\bf{-k-q}}}_{\downarrow}\rangle$. These finite momentum scattering leads to additional van Hove singularities in the spectral function which shows up as spin split features in the single particle DOS.  

\begin{figure*}[!t]
\centering
\includegraphics[width=15.5cm]{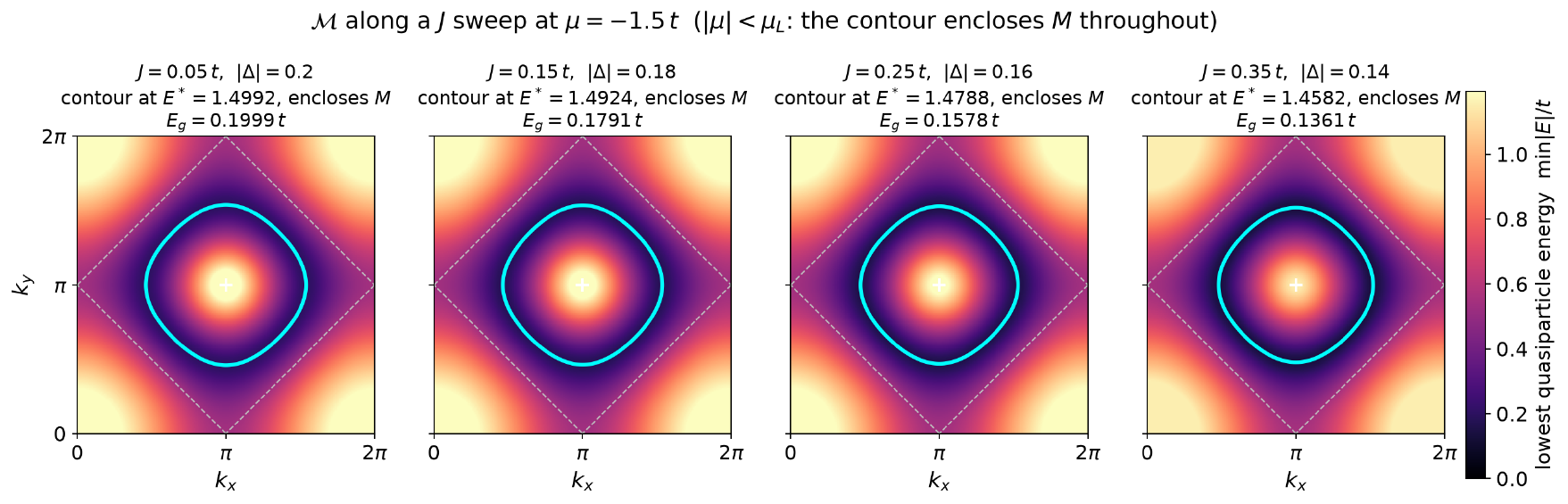}\\[-1mm]
\includegraphics[width=15.5cm]{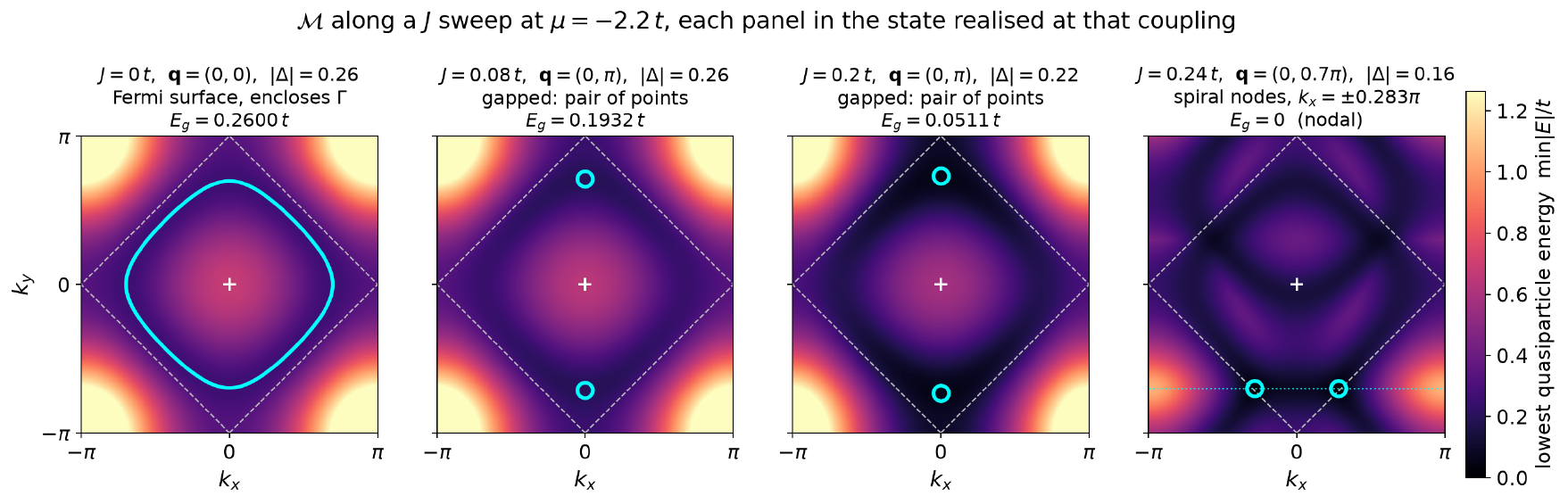}
\caption{The top panel encodes the fate of the minimum energy contour at constant $\mu=-1.5t$ as a function of increasing $J$. We see that for this value of the chemical potential, the 
minimum energy contour is always a one-dimensional object that encloses the $M$ point very much in analogy with the $J=0$ case. Thus the dimension of the minimum energy surface $\mathcal{M}$ is always fixed to one. The minimum energy locus for the case $\mu=-2.2t$ for various $J$ values. Here, one sees the Lifshitz-type transition where the dimensionality of the minimum energy locus is tuned from $\mathcal{M}=2$, enclosing the $\Gamma$ point to nodal points (corresponding to stripe order of the core spins), where the dimension of  $\mathcal{M}=0$ but with the spectral gap finite, and finally to the spirally ordered $(0,q)$ state with dimension of the minimum energy surface is $\mathcal{M}=0$ but with the spectral gap collapsed to zero, indicating a gapless superconductor.}
\label{fig:Jsweep}
\end{figure*}

We now turn our attention to the discussion of tuning of the minimum energy surface $\mathcal{M}$ in the case $J\neq 0$. In effect, there are two potential routes that one adopts: We start our discussion with the first case where $\mathcal{M}$ is tuned via the Kondo coupling $J$ whilst keeping $\mu$ fixed. The alternate route of sitting at a constant Kondo coupling and tuning $\mu$ will be discussed in Sec.~\ref{subsec:LifJmuneq0}.

Let us consider the case $\mu=-1.5t$ as a prototypical case of a scenario where the ordering wave vector is $(\pi,\pi)$. The discussions in Sec.~\ref{subsec:gapclpipi}, clearly attest to the fact that for this ferrimagnetically ordered superconducting state the gap closes if and only if $\Lambda =0$, which only entails for the flat band (at $\mu=0$), (we will get back to this case later in this section). Thus, for the case of the $(\pi,\pi)$ ordered superconducting state with increasing $J$, the system is fully gapped as evidenced by the spectral function and the lowest lying eigenvalues of the full BdG spectrum shown in the Fig.~\ref{fig5}: The figures clearly highlight that in the superconducting state, the ferrimagnetic ordering does not lead to a gapless phase and the minimum energy contour is always $\mathcal{M}=1$.

In contrast, from the phase diagram displayed in Fig.~\ref{fig2}, it is clearly seen that a different situation obtains at $\mu=-2.2t$:  The spectral function as well as the behavior of the lowest lying eigenvalues depicted in Fig.~\ref{fig:spec2.2}(a1)-(d1), and Fig.~\ref{fig:spec2.2}(a2)-(d2) clearly show that as the Kondo coupling is increased, the spectral gap in the superconductor decreases as the system transitions from the ferromagnet to the stripe ordered phase and finally vanishes for the $(0,q)$ ordered phase, (see Sec~\ref{sec:PD}, for a detailed discussion of the phase diagram). The above is completely in line with the results encapsulated in Table~\ref{tab:bzsummary-long}, where for $\Lambda\neq0$, the $(0,\pi)$ stripe-ordered superconductor have their spectral minima at two points (which are still gapped). Furthermore, as the superconducting system is tuned into the spiral at $(0,q)$ state, a gapless state conforming to Eq.~\ref{eq:L1} is formed that for $\mu=-2.2t$, lives on the locus $\mathcal{L}_1$ of Eq.~\ref{eq:lines}. Thus $\mathcal{M}$ transitions from $\mathcal{M}:1\rightarrow 0$.

The contrast between the situation that entails between the $\mu=-1.5t$ and $\mu=-2.2t$ couched in the language of the Lifshitz phenomenology is highlighted in Fig.~\ref{fig:Jsweep}. For the case $\mu=-1.5t$, the minimum energy surface related to the $(\pi,\pi)$ ordering wave vector satisfies the equation:
\begin{equation}
  \cos k_x + \cos k_y \;=\; \frac{1}{2t^2}\left(\mu^2 - \frac{h^2{(\mu^2+\Delta^2)}^2}{\mu^2}\right) - 2,
  \label{eq:ferricontour}
\end{equation}
The above equation is structurally a renormalized form of Eq.~\ref{eq:min_condn}. In other words, Eq.~\ref{eq:ferricontour} is very reminiscent of 
the $J=0$ case wherein the chemical potential is shifted (renormalized) via the shift $\mu_{\rm eff}\equiv\mu^2 \rightarrow \mu^2- h^2E_B^2/\mu^2$. In a similar vein, the spectral gap $E_{\bf min}$ conforms to the form: 
\begin{equation}
  E_{\bf min} \;=\; |\Delta|\sqrt{1 - \frac{h^2}{\mu^2}}
         \;=\; \frac{|\Delta|}{|\mu|}\sqrt{\mu^2 - h^2}
  \label{eq:ferriEmin}
\end{equation}
which is once again a simple field induced multiplicative renormalization of $\Delta$. The two equations Eq.~\ref{eq:ferricontour} and Eq.~\ref{eq:ferriEmin} contain within them the complete physics needed to understand Fig.~\ref{fig:Jsweep}, for $\mu=-1.5t$: The structure of the above two equations immediately implies that the Lifshitz type transition occurs when $\mu_{\rm eff}^{\rm L}=-2t$. Thus, at $\mu=-1.5t$, the minimal energy surface will always encircle the $M$ point, and the system does not undergo a Lifshitz transition as a function of $J$. In contrast the behavior is much richer for $\mu=-2.2t$. Here $|\mu|>\mu_{\rm eff}^{\rm L}$, and one finds that the minimal energy surface encloses the $\Gamma$ point. As one transitions into the superconductor with $(0,\pi)$ order, two gapped minima with spectral gap $E_g\neq 0$ are obtained, consistent with the results shown for $(0,\pi)$ order (entry III of Table~\ref{tab:bzsummary-long}, with $\Lambda\neq 0$).
Thus, for the ferrimagnetically ordered core spins, for $\mu\neq 0$, $\mathcal{M}$ is a one dimensional contour and the minimum energy that entails from the BdG spectrum $E_{\bm min} \neq 0$. This scenario is represented very well in Fig.~\ref{fig:Jsweep} by the case $J=0.08t$, and $J=0.20t$. 

As one evolves into the spirally ordered $(0,q)$ phase for the core spins, one finds that, consistent with the gap-closing condition of Eq.~\ref{eq:L1}, two gapless nodes are produced 
lying on the locus ${\cal L}_1$, consistent with the chemical potential window given by Eq.~\ref{eq:window}. At $\mu=0$, the flat band sits exactly at the Fermi level. According to the phase diagram elucidated in Fig.~\ref{fig2}at $\mu=0$, the preferred ordering wave vector is $(\pi,\pi)$. 
This corresponds to the case $\Lambda=0$ (entry II of Table~\ref{tab:bzsummary-long}), which states that the entire zone could go gapless. However, as seen in our discussion in Sec.~\ref{sec:PD}, such a gapless phase is not thermodynamically stable. Thus, we do not discuss this here in more detail.

Rather here, we quickly turn our attention to the elucidation of the generic behavior of the spectral function and the lowest eigenvalues (plotted in Fig.~\ref{fig7}(a1)-(d1), and Fig.~\ref{fig7}(a2)-(d2) respectively). As seen in these figures,  at this value of the chemical potential, the minimum energy is always locked to $|\Delta-h|$, independent of $\kv$. Tuning $J$ we have reiterated above as does not lead to the closing of the spectral gap and thus $\Mset=2$ always independent of $J$, (inclusive of $J=0$). This implies that there is no $J$-dependent renormalization of the minimum energy surface at all. The only effect of the finite $J$ coupling is seen in the Zeeman splitting of the flat-band.

\subsection{The Behavior of the DoS}
\label{subsec:Dos_Jneq0}
We now turn our attention to the analysis of the DoS for various chemical potential as a function of the Kondo-coupling $J$. We shall first discuss the cases at $\mu=-1.5t$ and $\mu=2.2t$, where the dispersive bands are involved, and finally turn to the case at $\mu=0$.

Concomitant to Sec.~\ref{sec:J0}, where we analyzed the behavior of the DoS for the case $J=0$, here we turn our attention to the behavior of the the DoS for the case where the Kondo coupling is at a finite non-zero value. To do so, let us look at the case where the chemical potential is fixed at $\mu=-1.5t$. As discussed in Sec.~\ref{sec:PD}, the core spins are aligned in the $q\equiv(\pi,\pi)$ ordered state. Furthermore, as seen in Sec.~\ref{subsec:spec_Jneq0}, at this value of the chemical potential the minimum energy contour is always $\Mset=1$, and the superconducting gap never closes. The clean signature of the $(\pi,\pi)$ ordered state can be seen in the behavior of the flat-bands under the influence of the Kondo coupling: For a finite J, as seen in Fig.~\ref{fig5}, the flat band splits and gives four $\delta$ function peaks at $\pm(E_B-h)$, and $\pm(E_B+h)$, (with $E_{\bm B} =\sqrt{\mu^2+\Delta^2}$). In other words the splitting in the flat-band contribution to the DoS is precisely pinned to the value $2J$ which is clearly seen in the DoS. Furthermore, as compared to the case at the $J=0$, for finite the peak position in the flat-band contribution to the DoS is halved by a factor of $1/2$.

The above situation for $\mu=-1.5t$ is in contrast with what is seen in the DoS for the case of $\mu=-2.2t$. In this case, as seen in Sec.~\ref{sec:PD}, the predominant order is the $(0,\pi)$ order which then goes into the spiral $(0,q)$. The $(0,q)$ order supports the gapless phase seen in Sec.~\ref{subsec:spec_Jneq0}, (see Fig.~\ref{fig:spec2.2}(d1), and Fig.~\ref{fig:spec2.2}(d2)). The form of the flat band contribution to the DoS is very indicative of the ordered state that obtains at a particular value of $J$ at a fixed $\mu$. For the case where commensurate orders are preferred, in the band basis the system feels an exchange field given by $hM_{\ell,\ell^\prime}$, which is purely diagonal. In contrast, for the stripe order $(0,\pi)$, and $(0,q)$ the exchange field is no longer purely diagonal, which means there is a mixing between the bands. In particular, this implies, a mixing between the flat and dispersive band which is an $O(J^2)$ effect. In other words the distribution of the overlap function $|M_0|$  in the case of the $(0,q)$ and $(0,\pi)$ orders becomes momentum dependent, and varies across the Brillouin zone. Thus, the DoS reflects the distribution of $|M_0|$ broadened by $J$ around $E_B=\sqrt{\mu^2+\Delta^2}$. This implies that the peaks are pinned at $E_B\pm h$. However, the width grows monotonically with $q$. Thus, the splitting becoming resolved only when $q$ is small. A fact attested to when one compares Fig.~\ref{dosjneq0}(d2), to Fig.~\ref{dosjneq0}(b2) and Fig.~\ref{dosjneq0}(c2): For the case with $J=0.24$ with $\qv=(0,0.7\pi)$, the $\delta$ function contribution to the DoS stemming from the flat band counterparts are much better resolved than their counterparts at $(0,\pi)$   

\begin{figure*}[!t]
\centering
\includegraphics[width=15.5cm]{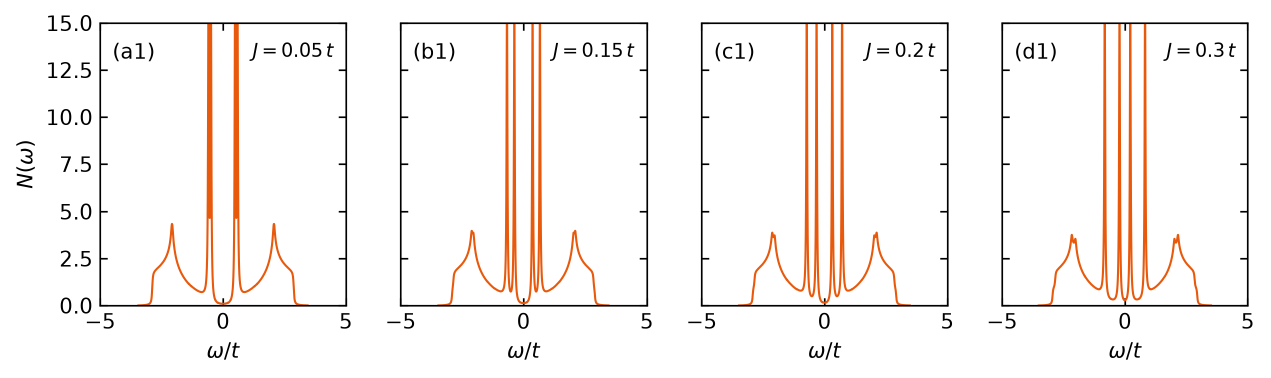}\\[-1mm]
\caption{Single particle DOS at selected $J/t$ showing the impact of the
Kondo interaction at $\mu=0t$. As elucidated in Sec.~\ref{subsec:Dos_Jneq0}, for $\mu=0$, the DoS ((see panels (a1)-(d1))clearly shows the effect of the predominant $(\pi,\pi)$ order 
that obtains here for finite $J$ wherein the system is always gapped and the flat-bands effectively gives Zeeman split $\delta$-functions with the splitting exactly equal to $2J$.Due 
to the fact that the chemical potential lies on a flat-band, the problem reduces to the problem of a $2\times 2$ system in a Zeeman field, leading to exact $\delta$ function contributions to the  
DoS at $\omega =\pm J, \mp J$. See, text for further details.}
\label{fig8}
\end{figure*}

Now that we have discussed the case at $\mu=1.5t$, and $\mu=-2.2t$, where predominantly the lower band is involved, let us understand the behavior of DoS for the case $\mu=0$, (half filled lattice), where the Fermi-level sits exactly at the flat-band, see Fig.~\ref{fig8}. As discussed earlier, the core spins at this chemical potential conform to $(\pi,\pi)$ order. This in turn implies that the flat bands feel an effective field given by $-J$, and the flat band decouples exactly from the dispersive pair, and the system reduces exactly to a two (flat)-level problem in a Zeeman field. This gives an eigenvalue spectrum $(\pm J,\mp J)$. The impact of the spectrum can be seen in the DoS, the square-root singularity that one usually associates with the superconducting DoS at the band edge is now replaced by the $\delta$-function contribution that are indicative of the fact that the flat-bands that are involved are dispersionless, (independent of $\kv$). The  splitting of the delta functions is percisely locked at $2J$, (refer Fig.~\ref{fig8}).The $\delta$ function sits on top of and overwhelms a step-function behavior that is predicated on the dispersive band eigenvalue having a minimum exactly at the $M$ point at $(\pi,\pi)$, (where it touches the flat-band). The rest of the features of the DoS can be attributed to a logarithmic peak that obtains via a saddle, and a step-down discontinuity due to a band edge.

\begin{figure*}[t]
\centering
\includegraphics[width=\textwidth]{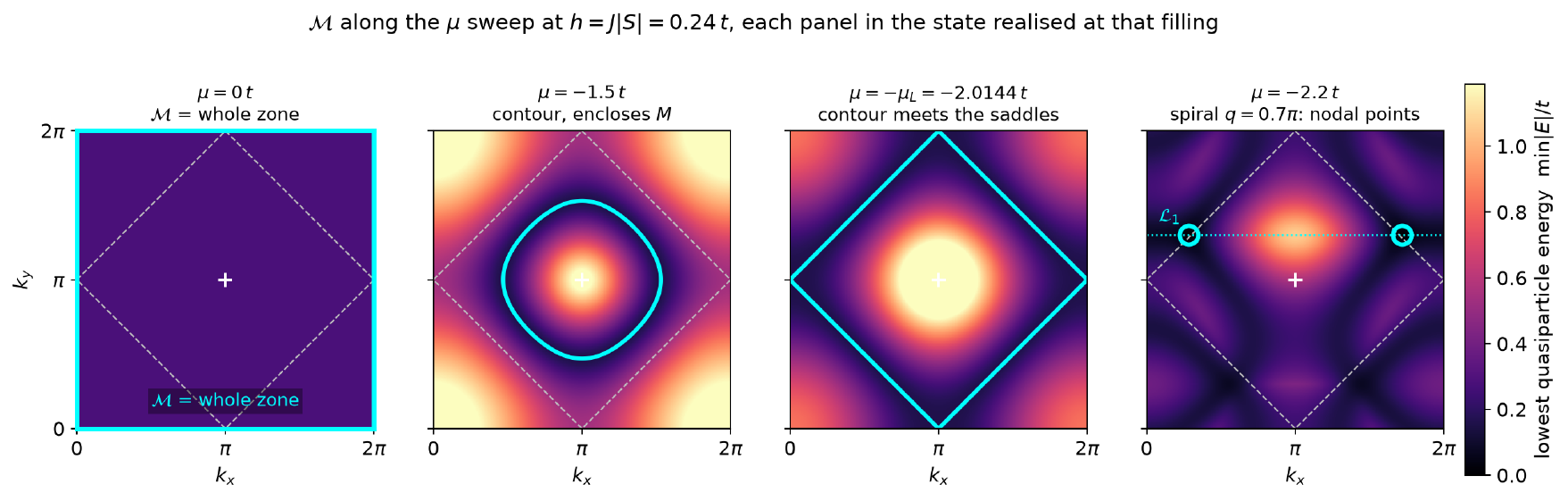}
\caption{Locus of gap minima, 
at a fixed $J$ as a function of $\mu$: For $|\mu|< -2t$, the minimum energy contour encircles the $M$ point, as the $\mu$ is decreased 
at a critical value $\mu_{\rm eff}=-2t$ the contour is the square
$|k_x-\pi|+|k_y-\pi|=\pi$, touching the zone boundary at $X$. Finally, a gapless nodal scenario obtains for $\mu=-2.2t$.
}
\label{fig:Lifshitzjneq0}
\end{figure*}

\subsection{Lifshitz Transition via the Chemical Potential}
\label{subsec:LifJmuneq0}
The behavior of the spectral function (and the DoS) above has revealed a Lifshitz-type transition that entails, as one sits at a fixed chemical potential and tunes the transition via the 
Kondo coupling $J$. However, experimentally it is very difficult to tune $J$; rather, it is easier to create samples with the value of $J$ fixed and tune $\mu$, for instance via doping. In this section we briefly describe the Lifshitz transition that entails if we follow this route.

This sequence leading to the change in the geometry of the minimum energy surface while tuning $\mu$ (at fixed J=h=0.24) is shown in Fig.~\ref{fig:Lifshitzjneq0}. The structure of the minimum energy surface for the case of the $(\pi,\pi)$ ordered state looks very similar to the case $J=0$ can be understood from the form of Eq.~\ref{eq:ferricontour} in comparison with the equivalent form of the minimum energy contour for the case $J=0$, see Eq.~\ref{eq:min_condn}. As we have discussed before, this clearly implies a simple multiplicative renormalization of the chemical potential $\mu_{\rm eff}$ due to the Kondo term, (see discussion centred around Eq.~\ref{eq:ferricontour}). In a similar vein, the gap itself gets multiplicatively renormalized, and we get 
$\vert \Delta\vert\rightarrow \vert \Delta\vert\sqrt{1-h^2/\mu^2}$.These two guarantees that the minimum energy surfaces look very identical. Of course, the value of the chemical potential where the Lifshitz-style transition occurs is now given by $\mu_{\rm eff}^{\rm L}=-2t$. The minimal energy surface at the transition $\mu_{\rm eff}^{\rm L}=-2t$ is given by the same equation as the $J=0$ case, namely $|k_x-\pi|+|k_y-\pi|=\pi$ (drawn as a reference contour in all the panels of Fig.~\ref{fig:Lifshitzjneq0}). The salient difference with the $J=0$ case occurs when we tune the system into the spirally ordered state at $\mu=-2.2$: The figure clearly shows the evolution into nodal points as the chemical potential is tuned to the gap closing nodal points. As shown in Fig.~\ref{dosjneq0}(d2), this leads to the gapless DoS that has deep consequences for observable quantities such as the specific heat: The specific heat, for instance, becomes $C\sim T^2$ as opposed to being exponentially suppressed. In a similar vein the ratio of the thermal conductivity to temperature $\kappa/T$ \cite{Durst2000,graf1996} is a constant for the case $J=0.24$ at $\mu=-2.2$, as opposed to exponentially suppressed for situations where an actual superconducting gap persists.
\section{Conclusions}
\label{sec:conc}
Research on flat band systems with multiple electronic bands is currently in vogue, both from the experimental and theoretical perspectives. More often than not, these systems are found to break free of the conventional norms of our understanding of strong correlation physics. In this paper, we have brought to the fore two types of Lifshitz type transitions in multi-band superconductors. Our model of choice is was the attractive Hubbard model defined on a Lieb lattice supplanted by a Kondo term that couples classical core spins to the itinerant fermions. Just like in the square-lattice implementation of the model \cite{mk_prb2016}, we saw that the interplay of the magnetic and superconducting fluctuations led to a very rich phase diagram with coexisting magnetic and superconducting orders. In particular, for incommensurate spiral $(0,q)$ order one found that the system hosts a phase where the spectral gap has essentially collapsed to zero nevertheless  the system still as evidenced by the structure factor, supports superconducting order.

Apart from the question of the ground state phase diagram that entails as a function of the interplay of the magnetic and superconducting fluctuations, the paper also was concerned with studying Lifshitz type transitions in this system.  For the case where the coupling to the core spins are turned off, one obtains a $\mu$ driven Lifshitz type transition. In fact tuning $\mu$ one obtains a minimum energy contour changes topology from an electron-like behavior to a hole-like behavior. This change in the connecting topology of the minimum energy surface can be, for instance, seen in Hall measurements.

For the case $J\neq 0$, the Lifshitz like transition can be obtained via following two distinct paths:$(a)$ By sitting at a fixed chemical potential and tuning $J$, and $(b)$ the more experimentally appealing alternative of sitting at a constant $J$ and tuning $\mu$. In either of the cases one clearly sees that the dimensionality of the minimum energy surface $\Mset$could be tuned from $1\rightarrow 0$. To show this evolution of $\Mset$, the equation for the contour of minimum energy was derived explicitly for each of the ordered states that the core spins obtain. Furthermore, through this analysis the gap closing condition was derived and it was explicitly shown that concomitant to the discussion in the derivation of the phase diagram, (see Sec.~\ref{sec:PD}), only the spiral states support a state with $E_g=0$. We have also shown that the thermodynamic quantities like the specific heat and transport properties such as the thermal conductivity are sensitive to the gap closing.

Experimentally, the Lieb lattice is relatively rare, and most of its experimental implementations are from synthetic platforms such as the optical lattice setup \cite{xing_prb2010,manninen_pra2010,chien_jphysb2016,takahashi_sciadv2015,takahashi_natcom2020}. Another approach is to engineer Lieb-lattices at the nano-scale using atomic manipulation through scanning probe microscopy \cite{drost_natphys2017,slot_natphys2017}. Photonic 
lattices and waveguides \cite{baba_natphot2008,yoshino_jpcm2004,kajiwara_prb2016,xia_opticlett2016,mukherjee_prl2015} provide another avenue to realize such lattices. More recently, Lieb lattice has come up as a promising candidate in the context of organic polymers viz. conjugated polymers, metal-organic frameworks (MOF) and covalent-organic frameworks (COFs) \cite{jiang_accchemres2021,ni_mathorizon2022,wang_chemsocrev2021,cui_topo2020}. These chemically tuneable 
porous systems, with their wide selection of substrates, chemical composition, and structural topologies, allow for high degree of tunability, making them an ideal platform for 2D functional materials \cite{liu_jamchmesoc2020,yuan_jamchemsoc2020,yuan_natcom2018,oi_advphysres2025,cui_natcom2020,jiang_natcom2019}.

\section{Acknowledgements:} SP would like to acknowledge  the use of the high-performance computing facility (AQUA) at the Indian 
Institute of Technology, Madras, India. MK and RN acknowledge the support from the Department of Science and Technology, Govt. of India 
through the grant CRG/2023/002593. RN acknowledges funding from the Center for Quantum Information Theory in Matter and Space-time, IIT Madras, and from the Department of Science and Technology, Government of India, under Grant No. DST/ICPS/QuST/Theme-3/2019/Q69, as well as support from the Mphasis F1 Foundation via the Centre for Quantum Information, Communication, and Computing (CQuICC). AI tools such as Claude and ChatGPT were used for doing certain algebraic manipulations, numerical checks and for writing the momentum space codes.
\appendix
\section{Uniform-order formalism}
\label{app:action}
The Appendix is dedicated to the spectral properties of Eq. (2) at various spin configurations and respective nodal conditions via mean-field BdG-type analysis via band-basis formalism, where individual band and interband contributions are easy to incorporate, hence suitable flat bands. It also contains the analytical forms for the gap closing conditions at the various wave vectors.
\subsection{Coherent-state action}
Our starting point is the partition function written in terms of fermionic coherent state path integral $Z=\int D[\bar c,c]\,e^{-S}$, with the action $S$ is given by:
\begin{equation}
S=\int_{0}^{\beta}\!d\tau\,\Big[\sum_{i\sigma}\bar c_{i\sigma}
(\partial_{\tau}-\mu)c_{i\sigma}+H[\bar c,c]\Big],
\label{eq:action}
\end{equation}
where $H$ is the Hamiltonian of Eq.~\eqref{eq:Ham} and the classical core
spins $\vec S_i$ is limited to lie in the $xy$ plane with wave vector, $\mathbf{q}$ as discussed in the main paper. Performing the Hubbard--Stratonovich  decoupling of the attractive paring term 
in Eq.~\ref{eq:action} we obtain:
\begin{equation}
H_{\Delta}=\sum_{i}\Big[\frac{\vert\Delta_{i}\vert^{2}}{\vert U\vert}
+\bar\Delta_{i}c_{i\downarrow}c_{i\uparrow}
+\Delta_{i}\bar c_{i\uparrow}\bar c_{i\downarrow}\Big].
\label{eq:HS}
\end{equation}
The saddle point of the
resulting action with respect to $\Delta_{i}$ reproduces the mean-field
Hamiltonian quoted in the main text, with $\Delta_{i}=-\vert U\vert\langle c_{i\downarrow}c_{i\uparrow}\rangle$.

\subsection{Uniform order parameters}
For the pairing, we take a uniform amplitude with the global phase
locked, such that $\Delta_{i}=\vert\Delta\vert$ real and site independent. Consistent with our assumption in the main part 
of the manuscript, we assume the core spins follow the planar spiral
$\vec S_i=(\cos\mathbf q\cdot\mathbf r_i,\sin\mathbf q\cdot\mathbf r_i,0)$,
where $\mathbf r_i$ is the position of site $i$, with no additional sublattice phase or offsets. The Kondo exchange term becomes
\begin{equation}
-h\sum_i\bigl(e^{-i\mathbf q\cdot\mathbf r_i}\bar c_{i\uparrow}c_{i\downarrow}
+e^{i\mathbf q\cdot\mathbf r_i}\bar c_{i\downarrow}c_{i\uparrow}\bigr),
\label{eq:exchange}
\end{equation}
with exchange field $h=J\vert\vec S_i\vert$, which specializes to $h=J$ for the case $S=1$. Crystal momenta $\mathbf k$ are measured in units of the inverse Lieb lattice constant, which spans two nearest-neighbor distances. The exchange term therefore transfers momentum $2\mathbf q$, coupling $\vert\mathbf k\uparrow\rangle$ to $\vert\kpq,\downarrow\rangle$.

\subsection{Bogoliubov--de Gennes kernel}
\label{app:kernel}
In the sublattice basis $(A,B,C)$ the noninteracting Bloch Hamiltonian is
\begin{equation}
H_0(\mathbf k)=\begin{pmatrix}0&-a&-b\\-a&0&0\\-b&0&0\end{pmatrix}
\label{eq:h0}
\end{equation}
with $t>0$ and the short-form notation:
\begin{equation}
a=2t\cos\frac{k_x}{2},\qquad b=2t\cos\frac{k_y}{2}.
\label{eq:hoppingdefs}
\end{equation}
For notational simplicity we denote quantities evaluated $\kpq$, e.g.,
$a'=2t\cos(k_x/2+q_x)$ and $b'=2t\cos(k_y/2+q_y)$ with primes. Furthermore again for notational ease, we collect the hoppings
into the vectors $\mathbf v=(a,b)$ and $\mathbf v'=(a',b')$, and use
$\mathbf v\cdot\mathbf v'=aa'+bb'$ and $\mathbf v\times\mathbf v'=ab'-ba'$.
The bands are $\epsilon^0_{\mathbf k}=0$ and
$\epsilon^\pm_{\mathbf k}=\pm E_{\mathbf k}$, with
$E_{\mathbf k}=\vert\mathbf v\vert$ (so $E_{\kpq}=\vert\mathbf v'\vert$),
and their real eigenvectors are
\begin{equation}
\vert u_0\rangle=\frac{1}{E_{\mathbf k}}
\begin{pmatrix}0\\-b\\a\end{pmatrix},\quad
\vert u_\pm\rangle=\frac{1}{\sqrt2E_{\mathbf k}}
\begin{pmatrix}E_{\mathbf k}\\\mp a\\\mp b\end{pmatrix}.
\label{eq:evecs}
\end{equation}
We write $\xi^\ell_{\mathbf k}=\epsilon^\ell_{\mathbf k}-\mu$ with $\ell=0,\pm$.

The Hamiltonian written in the spinor Nambu basis $\Psi_{\mathbf k}=(c_{\mathbf k\uparrow},c_{\kpq,\downarrow},
\bar c_{-\mathbf k\downarrow},\bar c_{-\mathbf k-2\mathbf q,\uparrow})$,
in which each entry has three sublattice components, the $12\times12$
kernel is
\begin{align}
{\cal H}&=
\begin{pmatrix}{\cal H}_0&\lvert\Delta\rvert\Sigma_z\\
\lvert\Delta\rvert\Sigma_z&-{\cal H}_0\end{pmatrix},\nonumber\\
{\cal H}_0&=\begin{pmatrix}H_0-\mu&-h\\-h&H_0'-\mu\end{pmatrix}.
\label{eq:H11}
\end{align}
Here $H_0'=H_0(\kpq)$, scalars multiply the $3\times3$ identity $I_3$,
and $\Sigma_z=\sigma^z\otimes \mathbb{I}_3$, with $\sigma^z$ acting on the two
momentum--spin sectors. We take $h\geq0$ and $\lvert\Delta\rvert>0$,
except in the normal-state rows of Table~\ref{tab:bzsummary-long}.

\subsection{Determinant and node condition}
\label{subsec:criterion}
Using the Pauli matrices $\tau_{x,z}$ that operates in the particle--hole sector, the Hamiltonian Eq.~\ref{eq:H11} is recast in the tensorial form as :${\cal H}=\tau_z\otimes{\cal H}_0+\tau_x\otimes\lvert\Delta\rvert\Sigma_z$.
Since $\Sigma_z{\cal H}_0$ is real,
\begin{equation}
\det{\cal H}=\bigl\vert\det(\Sigma_z{\cal H}_0-i\lvert\Delta\rvert I_6)\bigr\vert^2
\equiv R^2+I^2,
\label{eq:detsplit}
\end{equation}
where $R$ and $I$ are the real and imaginary parts of the determinant. The 
zero mode condition requires $R=I=0$ simultaneously. For later use, we define:
\begin{equation}
\Lambda\equiv h^2-\mu^2-\lvert\Delta\rvert^2,
\label{eq:Lambda}
\end{equation}
which vanishes at $h^2=\mu^2+\lvert\Delta\rvert^2$, (${\bf k}$-independent), exceptional value. The imaginary and real parts can be derived respectively to be:
\begin{equation}
I=2\lvert\Delta\rvert\mu\Lambda\,(\mathbf v^2-\mathbf v'^2),
\label{eq:Igeneral}
\end{equation}
\begin{align}
R={}&\Lambda\bigl[\Lambda^2+(\mathbf v\cdot\mathbf v')^2
-2h^2\,\mathbf v\cdot\mathbf v'\nonumber\\
&-(\mu^2-\lvert\Delta\rvert^2)(\mathbf v^2+\mathbf v'^2)\bigr]\nonumber\\
&-(\mu^2+\lvert\Delta\rvert^2)(\mathbf v\times\mathbf v')^2.
\label{eq:Rgeneral}
\end{align}
For $\mu\neq0$ and $\Lambda\neq0$, a node therefore requires
$\vert\mathbf v\vert=\vert\mathbf v'\vert$ and $R=0$, and the nodes are isolated points wherever these constraints are locally independent. At $\mu=0$,
$I\equiv0$. At $\Lambda=0$, the condition reduces to
$\mathbf v\times\mathbf v'=0$. These exceptional cases can support nodal
curves or a zero set covering the whole Brillouin zone (BZ); see
Table~\ref{tab:bzsummary-long}. We caution that all conditions derived in these Appendixes are purely spectral, especially as pertains to the behavior of the closing of the quasi-particle excitation gap. Whether such states exhibit gapless superconductivity for a particular ordering wave-vector is a question of thermodynamic stability which requires a separate free-energy comparison.

\subsection{Band basis}
\label{subsec:bandbasis}
Rotating each sublattice triplet to the band basis of Eq.~\eqref{eq:evecs},
$c_{\mathbf k\ell\sigma}=\sum_\alpha\langle u_\ell(\mathbf k)\vert\alpha\rangle
c_{\mathbf k\alpha\sigma}$, leaves the pairing diagonal in the band index,
while the exchange mixes bands through the overlap
$M_{\ell\ell'}(\mathbf k)=\langle u_\ell(\mathbf k)\vert u_{\ell'}(\kpq)\rangle$.
In band order $(0,+,-)$,
\begin{align}
M_{00}&=\frac{\mathbf v\cdot\mathbf v'}{\vert\mathbf v\vert\vert\mathbf v'\vert},
\nonumber\\
M_{\pm0}&=-M_{0\pm}=\pm\frac{\mathbf v\times\mathbf v'}
{\sqrt2\,\vert\mathbf v\vert\vert\mathbf v'\vert},\nonumber\\
M_{\pm\pm}&=1-M_{\pm\mp}=\frac12+\frac{\mathbf v\cdot\mathbf v'}
{2\vert\mathbf v\vert\vert\mathbf v'\vert}.
\label{eq:Mgeneral}
\end{align}
We denote this general matrix by $M$. It requires
$\vert\mathbf v\vert,\vert\mathbf v'\vert>0$; The eigenbasis is not unique  at the  band-touching point 
whereas Eqs.~\eqref{eq:Igeneral} and \eqref{eq:Rgeneral} remain
well defined.

With $\hat\xi=\mathrm{diag}(\xi^0_{\mathbf k},\xi^+_{\mathbf k},\xi^-_{\mathbf k})$
and $\hat\xi'$ the same at $\kpq$, the band-basis kernel is (scalars
again multiply $I_3$)
\begin{equation}
{\cal H}_{\rm band}=\begin{pmatrix}
\hat\xi&-hM&\lvert\Delta\rvert&0\\
-hM^T&\hat\xi'&0&-\lvert\Delta\rvert\\
\lvert\Delta\rvert&0&-\hat\xi&hM\\
0&-\lvert\Delta\rvert&hM^T&-\hat\xi'
\end{pmatrix}.
\label{eq:bandkernel}
\end{equation}
When $M$ is a signed permutation, band $\ell$ at $\mathbf k$ couples only
to band $\ell'$ at $\kpq$, with $M_{\ell\ell'}=\pm1$, and
${\cal H}_{\rm band}$ splits into three $4\times4$ blocks with
\begin{align}
\det{\cal H}_{\ell\ell'}={}&\bigl(\xi^\ell_{\mathbf k}\xi^{\ell'}_{\kpq}
+\lvert\Delta\rvert^2-h^2\bigr)^2\nonumber\\
&+\lvert\Delta\rvert^2\bigl(\xi^\ell_{\mathbf k}-\xi^{\ell'}_{\kpq}\bigr)^2,
\label{eq:det4}
\end{align}
independent of the sign of $M_{\ell\ell'}$. A zero mode then requires
$\xi^\ell_{\mathbf k}=\xi^{\ell'}_{\kpq}$ and
$h^2=(\xi^\ell_{\mathbf k})^2+\lvert\Delta\rvert^2$. This reduction holds into the $4\times 4$ form holds throughout the BZ for the ferromagnet and the ferrimagnet, and only on special loci for the stripe and the spiral.

\section{Ferromagnet, $\mathbf q=(0,0)$}
\label{app:ferro}
In case of ferromagnet, with $q = (0,0)$, $\hat\xi'=\hat\xi$ and the overlap matrix $M$ is the identity,
\begin{equation}
M=\mathbb{I}_3.
\label{eq:Mferro}
\end{equation}
Each band forms an independent Zeeman-split superconducting block, with
quasiparticle energies
$E_{\ell,\pm}=\bigl|[(\xi^\ell_{\mathbf k})^2+\lvert\Delta\rvert^2]^{1/2}\pm h\bigr|$.
Equivalently, Eq.~\ref{eq:Igeneral} gives $I=0$ and Eq.~\ref{eq:Rgeneral} and Eq.~\ref{eq:det4} are
\begin{equation}
R=\prod_{\ell=0,\pm}\bigl[h^2-(\xi^\ell_{\mathbf k})^2-\lvert\Delta\rvert^2\bigr],
\qquad \det{\cal H}=R^2.
\label{eq:detferro}
\end{equation}

\subsection{Gap closing}
\label{subsec:gap_clq00}
From the above equations, Eq.~\ref{eq:detferro}, we get conditions for the zero modes to occur where
$(\xi^\ell_{\mathbf k})^2=h^2-\lvert\Delta\rvert^2$, which sets the spectral
threshold
\begin{equation}
h_c=\min_{\ell,\mathbf k}\bigl[(\xi^\ell_{\mathbf k})^2+\lvert\Delta\rvert^2\bigr]^{1/2}.
\label{eq:hcferro}
\end{equation}
The dispersive bands generally give contours, whereas the flat band closes
over the whole BZ, and only at $\Lambda=0$ (class~I of
Table~\ref{tab:bzsummary-long}). The above discussion is purely about the spectrum. However, whether a gapless phase is thermodynamically stable for the 
$(0,0)$ ordered state is a question about thermodynamic stability\cite{clogston,chandrasekhar,sarma1963,liuwilczek2003}.

\section{Ferrimagnet, $\mathbf q=(\pi,\pi)$}
\label{app:ferri}
For $q=(0,0)$, the state is symmetrically equivalent to an antiferromagnet in the Lieb lattice, producing $a'=-a$ and $b'=-b$. So $H_0'=-H_0$, $E_{\kpq}=E_{\mathbf k}$, as a result the overlap is the signed permutation
\begin{equation}
M=\mathbb{P}=\begin{pmatrix}-1&0&0\\0&0&1\\0&1&0\end{pmatrix},
\label{eq:ffpipi}
\end{equation}
indicating that the flat band stays unchanged, but the other two dispersive bands (denoted with $\pm$) interchange. The three blocks of Eq.~\eqref{eq:det4} are
\begin{equation}
\begin{array}{c | c | c | c}
(\ell,\ell') & \xi^\ell_{\mathbf k} & \xi^{\ell'}_{\kpq} & M_{\ell\ell'}\\
\hline
(0,0) & -\mu & -\mu & -1\\
\hline
(+,-) & E_{\mathbf k}-\mu & -E_{\mathbf k}-\mu & 1\\
\hline
(-,+) & -E_{\mathbf k}-\mu & E_{\mathbf k}-\mu & 1
\end{array}
\label{eq:pipiblocks}
\end{equation}

\subsection{Gap closing}
\label{subsec:gapclpipi}
Inserting Eq.~\eqref{eq:pipiblocks} into Eq.~\eqref{eq:det4} gives the conditions for the zero mode conditions as $I=0$ and
$\det{\cal H}=R^2$, with
\begin{equation}
R=\Lambda\bigl[(\Lambda+E_{\mathbf k}^2)^2+4\lvert\Delta\rvert^2E_{\mathbf k}^2\bigr].
\label{eq:detpipi}
\end{equation}
The first factor comes from the flat band and the second from the two
dispersive blocks. For $\lvert\Delta\rvert>0$ the spectrum is gapped unless
$\Lambda=0$, where the flat sector closes throughout the
BZ (class~II of Table~\ref{tab:bzsummary-long}). Whether this point lies
inside the stable ferrimagnetic phase is decided by the free-energy
calculation of the main text.

\section{Stripe, $\mathbf q=(0,\pi)$}
\label{app:0pi}
The stripe order along the y-direction results in the ordering vector $q=(0,\pi)$, giving $\mathbf v'=(a,-b)$, so $\mathbf v'^2=\mathbf v^2=a^2+b^2$,
$\mathbf v\cdot\mathbf v'=a^2-b^2$, $\mathbf v\times\mathbf v'=-2ab$, and
Eq.~\eqref{eq:Mgeneral} becomes
\begin{equation}
M=\mathbb{S}(a,b)=\frac{1}{a^2+b^2}
\begin{pmatrix}
a^2-b^2&\sqrt2ab&-\sqrt2ab\\
-\sqrt2ab&a^2&b^2\\
\sqrt2ab&b^2&a^2
\end{pmatrix}.
\label{eq:ff0pi}
\end{equation}
The bands are generally mixed; $M$ reduces to $\mathbb{I}_3$ at $b=0$ and to $\mathbb{P}$ at $a=0$, away from the band-touching point. In other words, this simplifications obtains at $k_y=\pi$, where the matrix $M_{\ell\ell'}=\delta_{\ell\ell'}$ like the ferromagnetic case, or at $k_x=\pi$, where the $M_{\ell\ell'}$ is exactly the 
signed permutation matrix of Eq.~\ref{eq:ffpipi}.Since, $\vert\mathbf v'\vert=\vert\mathbf v\vert$, $I\equiv0$, and
Eq.~\eqref{eq:Rgeneral} gives the exact stripe zero-mode condition
\begin{align}
&\Lambda\bigl[\Lambda^2+(a^2-b^2)^2-2h^2(a^2-b^2)\nonumber\\
&\quad-2(\mu^2-\lvert\Delta\rvert^2)(a^2+b^2)\bigr]
=4(\mu^2+\lvert\Delta\rvert^2)a^2b^2.
\label{eq:Fstripe}
\end{align}
At $\Lambda=0$ it reduces to $ab=0$. The resulting loci
form class~III of Table~\ref{tab:bzsummary-long}. Once again, their spectral existence
does not imply stability in the phase diagram of Sec.~\ref{sec:PD}.

\section{Spiral, $\mathbf q=(0,q)$}
\label{app:0q}
For spiral phase with ordering vector with general $q$ where $0<q<\pi$ and the parameters, $I=0$ requires
$\vert\mathbf v\vert=\vert\mathbf v'\vert$. Since
\begin{align}
\mathbf v^2-\mathbf v'^2={}&4t^2[\sin q_x\sin(k_x+q_x)\nonumber\\
&+\sin q_y\sin(k_y+q_y)].
\label{eq:master}
\end{align}
Hence, the candidate loci are the lines
\begin{equation}
{\cal L}_1:\ k_y=-q,\qquad {\cal L}_2:\ k_y=\pi-q\pmod{2\pi}.
\label{eq:lines}
\end{equation}
On ${\cal L}_1$, $b'=b$ and $M=\mathbb{I}$, so the block decomposition is
exact. On ${\cal L}_2$, $b=-b'=2t\sin(q/2)$ and $M=\mathbb{S}(a,b)$; the bands generally remain coupled.

\subsection{Gap closing}
\label{subsec:gapclosing0q}
Since different conditions on parameters produce different loci of zero modes, firstly on ${\cal L}_1$, $R$ takes the ferromagnetic form of
Eq.~\eqref{eq:detferro}, and zero modes occur at
\begin{equation}
\cos^2\frac{k_x}{2}=\frac{\bigl(\mu\pm\sqrt{h^2-\lvert\Delta\rvert^2}\bigr)^2}{4t^2}
-\cos^2\frac q2,
\label{eq:L1}
\end{equation}
provided $h\geq\lvert\Delta\rvert$ and the right-hand side lies in $[0,1]$.
There are at most four such points, in two $\pm k_x$ pairs. On
${\cal L}_2$, the exact condition is Eq.~\eqref{eq:Fstripe} with
$b=2t\sin(q/2)$; it is quadratic in $a^2$ and gives at most four points.
Class~IV of Table~\ref{tab:bzsummary-long} lists the allowed roots and the
exceptional lines at $\Lambda=0$; the $\mu=0$ and other
limiting cases are collected in classes~V--VII.

Restricted to ${\cal L}_1$, the threshold
\begin{equation}
h_c^{({\cal L}_1)}=\min_{\ell,\,\mathbf k\in{\cal L}_1}
\bigl[(\xi^\ell_{\mathbf k})^2+\lvert\Delta\rvert^2\bigr]^{1/2}\geq\lvert\Delta\rvert
\label{eq:hc}
\end{equation}
reaches $\lvert\Delta\rvert$ when a dispersive band crosses the Fermi level
on this line, i.e., when
\begin{equation}
2t\lvert\cos(q/2)\rvert\leq|\mu|\leq2t\sqrt{1+\cos^2(q/2)}.
\label{eq:window}
\end{equation}
For a self-consistent solution satisfying Eq.~\eqref{eq:window} at the
onset, $h=J$ gives
\begin{equation}
J_g=|\Delta(J_g)|,
\label{eq:Jg}
\end{equation}
which is solved together with the mean-field equations.

\begingroup
\let\LiebOriginalIncludeGraphics\includegraphics
\renewcommand{\includegraphics}[2][]{%
  \IfFileExists{#2}{%
    \LiebOriginalIncludeGraphics[#1]{#2}%
  }{%
    \PackageWarning{lieb-kondo}{Missing schematic `#2'; using a placeholder}%
    \begingroup
      \setlength{\fboxsep}{0pt}%
      \fbox{\rule{0pt}{1.00cm}\rule{1.25cm}{0pt}}%
    \endgroup
  }%
}
\input{prb_zero_modes}
\endgroup

\bibliography{liebkondo}

\end{document}

%% file: prb_zero_modes.tex
\section{Summary of zero-mode loci}
\label{sec:bzsummary}

Table~\ref{tab:bzsummary-long} collects the zero-mode conditions and loci. 
The overlap
matrices are given in Eqs.~\eqref{eq:Mgeneral}, \eqref{eq:Mferro},
\eqref{eq:ffpipi}, and \eqref{eq:ff0pi}. ``Generic'' means
$\lvert\Delta\rvert>0$, $\mu\neq0$, and $\Lambda\neq0$, with $\Lambda$ from
Eq.~\eqref{eq:Lambda};
$R_{\mu=0}$ denotes Eq.~\eqref{eq:Rgeneral} evaluated at $\mu=0$.
Special-parameter rows take precedence over generic cases.
In the sketches, red marks indicate zero modes, dashed gray lines indicate
candidate loci, and pink shading indicates a zero set covering the whole BZ.
Curves and point positions are schematic; the straight nodal lines are
exact and drawn at $q=0.65\pi$. Opposite BZ edges are identified. A sketch
does not imply that roots exist for every choice of parameters.

\onecolumngrid
\begingroup
\small
\setlength{\tabcolsep}{2pt}
\setlength{\LTcapwidth}{\textwidth}
\setlength{\LTleft}{0pt}
\setlength{\LTright}{0pt}
\renewcommand{\arraystretch}{1.0}
\newcommand{\BZplaincell}[3]{%
  \parbox[c]{#1\textwidth}{#2\vspace{1pt}\strut #3\strut\par\vspace{1pt}}}
\newcommand{\BZplainrow}[6]{%
  \BZplaincell{.10}{\raggedright}{#1}&%
  \BZplaincell{.12}{\raggedright}{#2}&%
  \BZplaincell{.15}{\centering}{#3}&%
  \BZplaincell{.225}{\raggedright}{#4}&%
  \BZplaincell{.26}{\raggedright}{#5}&%
  \BZplaincell{.10}{\centering}{#6}}
\renewcommand{\BZrow}[6]{%
  \BZplainrow{#1}{#2}{#3}{#4}{#5}{\BZfigure{#6}}}
\begin{longtable}{@{}llllll@{}}
\caption{Zero-mode conditions, overlap matrices, and BZ loci, grouped by
magnetic order. Roman numerals label the seven phase or parameter classes;
numbered cases are local to each class. Sketches appear beside the locus
descriptions.}\label{tab:bzsummary-long}\\
\hline
\hline
\BZplainrow{\textbf{Phase / class}}{\textbf{Case}}{\textbf{Overlap $M$}}{\textbf{Zero-mode condition}}{\textbf{Locus}}{\textbf{BZ sketch}}\\
\hline
\endfirsthead
\multicolumn{6}{@{}l}{\textbf{Table \thetable\ (continued)}}\\[3pt]
\hline
\hline
\BZplainrow{\textbf{Phase / class}}{\textbf{Case}}{\textbf{Overlap $M$}}{\textbf{Zero-mode condition}}{\textbf{Locus}}{\textbf{BZ sketch}}\\
\hline
\endhead
\hline
\multicolumn{6}{r@{}}{\footnotesize Continued on the next page}\\
\endfoot
\hline
\hline
\endlastfoot
%
\BZrow{\textbf{I.}\par\smallskip \textbf{Ferro\-magnet}\par $(0,0)$}{\textbf{(1)}\enspace Dispersive bands}{$\mathbb{I}$}{$(\pm E_{\mathbf k}-\mu)^2=h^2-\lvert\Delta\rvert^2$,\par $h\geq\lvert\Delta\rvert$,\par $0\leq E_{\mathbf k}\leq2\sqrt2\,t$.}{\textbf{Contours (1D)} for allowed energies; they can shrink to points at band extrema or be absent.}{bz-contour}\\*
\cline{2-6}
\BZrow{}{\textbf{(2)}\enspace Flat band;\par $\Lambda\neq0$}{$\mathbb{I}$}{No flat-band solution.}{\textbf{Empty flat sector.} Dispersive zero modes are not excluded.}{bz-empty}\\*
\cline{2-6}
\BZrow{}{\textbf{(3)}\enspace Flat band;\par $\Lambda=0$}{$\mathbb{I}$}{Since $\xi^0_{\mathbf k}=-\mu$, Eq.~\eqref{eq:detferro} vanishes at every $\mathbf k$.}{\textbf{Whole BZ (2D).} The flat sector closes at every momentum.}{bz-whole}\\
\hline
%
\BZrow{\textbf{II.}\par\smallskip \textbf{Ferri\-magnet}\par $(\pi,\pi)$}{\textbf{(1)}\enspace $\Lambda\neq0$}{$\mathbb{P}$}{$R\neq0$ at every $\mathbf k$ [Eq.~\eqref{eq:detpipi}].}{\textbf{Empty.} The superconducting spectrum is gapped.}{bz-empty}\\*
\cline{2-6}
\BZrow{}{\textbf{(2)}\enspace $\Lambda=0$}{$\mathbb{P}$}{First (flat-band) factor of Eq.~\eqref{eq:detpipi} vanishes.}{\textbf{Whole BZ (2D)} through the flat sector. Additional degeneracy can occur at the band-touching point.}{bz-whole}\\
\hline
%
\BZrow{\textbf{III.}\par\smallskip \textbf{Stripe}\par $(0,\pi)$}{\textbf{(1)}\enspace $\Lambda\neq0$}{$\mathbb{S}(a,b)$}{Eq.~\eqref{eq:Fstripe}.}{\textbf{Contours (1D) at regular roots}, if physical roots exist. The curve is schematic.}{bz-curve}\\*
\cline{2-6}
\BZrow{}{\textbf{(2)}\enspace $\Lambda=0$,\par $b=0$}{$\mathbb{I}$}{$k_y=\pi\pmod{2\pi}$.}{\textbf{Horizontal line (1D).} The top and bottom edges are the same line on the BZ torus.}{bz-horizontal-edge}\\*
\cline{2-6}
\BZrow{}{\textbf{(3)}\enspace $\Lambda=0$,\par $a=0$}{$\mathbb{P}$}{$k_x=\pi\pmod{2\pi}$.}{\textbf{Vertical line (1D).} The left and right edges are the same line on the BZ torus.}{bz-vertical-edge}\\
\hline
%
\BZrow{\textbf{IV.}\par\smallskip \textbf{Spiral}\par $(0,q)$\par $0<q<\pi$}{\textbf{(1)}\enspace Generic;\par $\mathcal L_1$:\par $k_y=-q$}{$\mathbb{I}$}{$\begin{aligned}\cos^2\frac{k_x}{2}&=\frac{\bigl(\mu\pm\sqrt{h^2-\lvert\Delta\rvert^2}\bigr)^2}{4t^2}\\&\quad-\cos^2\frac q2.\end{aligned}$}{\textbf{At most four points (0D)}, in two $\pm k_x$ pairs. The right-hand side must lie in $[0,1]$. The dashed line is only a candidate locus.}{bz-l1-points}\\*
\cline{2-6}
\BZrow{}{\textbf{(2)}\enspace Generic;\par $\mathcal L_2$:\par $k_y=\pi-q$}{$\mathbb{S}(a,b)$}{Eq.~\eqref{eq:Fstripe} with\par $b=2t\sin(q/2)$,\par $a=2t\cos(k_x/2)$.}{\textbf{At most four points (0D)} from the physical roots of a quadratic in $a^2$. There may be no nodes.}{bz-l2-points}\\*
\cline{2-6}
\BZrow{}{\textbf{(3)}\enspace $\Lambda=0$,\par $k_y=-q$}{$\mathbb{I}$}{$b'=b$.}{\textbf{Entire horizontal line (1D)} $k_y=-q$. The candidate line becomes nodal at this exceptional field.}{bz-l1-line}\\*
\cline{2-6}
\BZrow{}{\textbf{(4)}\enspace $\Lambda=0$,\par $k_x=\pi$}{$\begin{cases}\mathbb{I},&bb'>0,\\\mathbb{P},&bb'<0.\end{cases}$}{$a=a'=0$.}{\textbf{Entire vertical line (1D)} $k_x=\pi$. Together with $k_y=-q$, this gives the complete exceptional-field spiral locus.}{bz-vertical-edge}\\
\hline
%
\BZrow{\textbf{V.}\par\smallskip \textbf{General order}\par $(q_x,q_y)$}{\textbf{(1)}\enspace Generic}{$\BZmatG$}{$\vert\mathbf v\vert=\vert\mathbf v'\vert$ \textbf{and} $R=0$.}{\textbf{Isolated points (0D)} where the two constraints are locally independent. Solutions need not exist.}{bz-points}\\*
\cline{2-6}
\BZrow{}{\textbf{(2)}\enspace $\Lambda=0$}{$\begin{cases}\mathbb{I},&\mathbf v\cdot\mathbf v'>0,\\\mathbb{P},&\mathbf v\cdot\mathbf v'<0\end{cases}$\par on nodes}{$\mathbf v\times\mathbf v'=0$.}{\textbf{Generally curves (1D)} where $\mathbf v$ and $\mathbf v'$ are collinear. Special orders can give the \textbf{whole BZ (2D)}.}{bz-curve}\\
\hline
%
\BZrow{\textbf{VI.}\par\smallskip \textbf{Any magnetic order}}{\textbf{(1)}\enspace $h<\lvert\Delta\rvert$}{Order-specific $M$}{Spectral gap $\geq\lvert\Delta\rvert-h>0$.}{\textbf{Empty.} There are no zero modes below the pairing scale.}{bz-empty}\\*
\cline{2-6}
\BZrow{}{\textbf{(2)}\enspace $h=\lvert\Delta\rvert$,\par $\mu\neq0$}{$\mathbb{I}$\par on nodes}{$a'=a$, $b'=b$,\par $E_{\mathbf k}^2=\mu^2$.}{\textbf{Points or contours}, depending on the order, or no solutions. The sketch shows the point case.}{bz-points}\\*
\cline{2-6}
\BZrow{}{\textbf{(3)}\enspace $\mu=0$,\par $h>\lvert\Delta\rvert$}{Generally $\BZmatG$}{$R_{\mu=0}=0$,\par $I\equiv0$.}{\textbf{Contours (1D) at regular roots}, if present. Equal dispersion $\vert\mathbf v\vert=\vert\mathbf v'\vert$ is not required.}{bz-curve}\\*
\cline{2-6}
\BZrow{}{\textbf{(4)}\enspace $\mu=0$,\par $h=\lvert\Delta\rvert$}{$\begin{cases}\mathbb{I},&\mathbf v\cdot\mathbf v'>0,\\\mathbb{P},&\mathbf v\cdot\mathbf v'<0\end{cases}$\par on nodes}{$\mathbf v\times\mathbf v'=0$.}{\textbf{Collinearity curves (1D)}, or the \textbf{whole BZ (2D)} for special orders.}{bz-curve}\\
\hline
%
\BZrow{\textbf{VII.}\par\smallskip \textbf{Normal state}\par $\lvert\Delta\rvert=0$}{\textbf{(1)}\enspace General parameters}{Order-specific $M$}{$\det\mathcal M=0$.}{\textbf{Fermi contours (1D)} in general, if present. Band edges and flat bands can change the dimension.}{bz-contour}\\*
\cline{2-6}
\BZrow{}{\textbf{(2)}\enspace $h=0$,\par $\mu=0$}{$hM=0$}{$\xi^0_{\mathbf k}=0$ at every $\mathbf k$.}{\textbf{Whole BZ (2D)} through the normal flat band.}{bz-whole}\\
\end{longtable}
\endgroup
\twocolumngrid

%% file: lieb_kondo_sc_RajModV21.bbl
\begin{thebibliography}{61}
\bibitem [{qso()}]{qsoftnote}%
  \BibitemOpen
  \href@noop {} {}\bibinfo {note} {The ordering wave vector is weakly selected
  and exhibits finite-size dependence. This is due to the fact that the
  classical spins couple only to the itinerant fermions. Thus, all $\bm q$ are
  degenerate at $h=0$, with the $\bm q$ dependence of the free energy entering
  at $\mathcal{O}(h^2)$. Thus, the selected $\bm q$ is in fact a function of
  system sizes. The values quoted in the text correspond to $L=20$ and should
  therefore be regarded as representative of a nearly degenerate
  manifold.}\BibitemShut {Stop}%
\bibitem [{unw()}]{unweighted}%
  \BibitemOpen
  \href@noop {} {}\bibinfo {note} {We are restricting ourselves to the
  unweighted DoS where the coherence factors are not taken into account. This
  produces results which are particle-hole symmetric.}\BibitemShut {Stop}%
\bibitem [{kgr()}]{kgrid}%
  \BibitemOpen
  \href@noop {} {}\bibinfo {note} {For a finite lattice of linear size $L$ with
  periodic boundary conditions, the allowed momenta are discretized in steps of
  $2\pi/L$.}\BibitemShut {Stop}%
\bibitem [{loc()}]{locus2}%
  \BibitemOpen
  \href@noop {} {}\bibinfo {note} {On the locus ${\cal L}_2$, the gap closing
  condition can be written as $a_2x^2+a_1x+a_0=0$, with $x\equiv
  4t^2\cos^2k_x/2$. The roots gives a real value for $k_x$ only if $0\le x \le
  4t^2$}\BibitemShut {NoStop}%
\end{thebibliography}
